\documentclass{aa}  

\usepackage{graphicx}
\usepackage{txfonts}
\usepackage{natbib}
\usepackage{comment}
\usepackage{amsmath}	% Advanced maths commands
\usepackage{amssymb}	% Extra maths symbols
\usepackage{subfigure}
\usepackage{enumerate}
\usepackage{csquotes}
\usepackage[dvipsnames]{xcolor}
\usepackage{gensymb}
\usepackage{mwe}
\usepackage{float}
\usepackage{soul}
\usepackage{comment}
\usepackage{rotating}
\usepackage{hhline}
\usepackage{booktabs}
\usepackage[pdfpagelabels=false]{hyperref}	% Hyperlinks

\catcode`\@=11
\def\gta{\ifmmode{\,\mathrel{\mathpalette\@versim>\,}}
    \else{$\,\mathrel{\mathpalette\@versim>}\,$}\fi}
\def\lta{\ifmmode{\,\mathrel{\mathpalette\@versim<\,}}
    \else{$\,\mathrel{\mathpalette\@versim<}\,$}\fi}
\def\@versim#1#2{\lower 2.9truept \vbox{\baselineskip 0pt \lineskip
    0.5truept \ialign{$\m@th#1\hfil##\hfil$\crcr#2\crcr\sim\crcr}}}
\catcode`\@=12  

\hypersetup{colorlinks=true,linkcolor=blue,citecolor=blue,filecolor=blue,urlcolor=blue,}

\newcommand{\kpc}   {\,{\rm kpc}}
\newcommand{\pc}    {\,{\rm pc}}

\newcommand{\Msun}  {\,{\rm M_{\odot}}}
\newcommand{\Gyr}   {\,{\rm Gyr}}

\begin{document}

   \title{Multiple stellar population mass loss in massive Galactic globular clusters}
   \titlerunning{Mass loss in massive Galactic GCs}
   \authorrunning{E. Lacchin et al.}

   %\subtitle{I. Overviewing the $\kappa$-mechanism}

   \author{E. Lacchin
          \inst{1,} \inst{2,} \inst{3,} \inst{4,} \inst{5,}
          A. Mastrobuono-Battisti \inst{3,6,7},
          F. Calura\inst{1},
          C. Nipoti\inst{2},
          A. P. Milone\inst{4,8},
          M. Meneghetti\inst{1},
          E. Vanzella\inst{1}.
}
   \institute{INAF - OAS, Osservatorio di Astrofisica e Scienza dello Spazio di Bologna, via Gobetti 93/3, I-40129 Bologna, Italy\\
    \email{elena.lacchin@inaf.it}
   \and
 Dipartimento di Fisica e Astronomia, Università di Bologna, via Gobetti 93/3, I-40129 Bologna, Italy
    \and
GEPI, Observatoire de Paris, PSL Research University, CNRS, Place Jules Janssen, 92195, Meudon,  France
   \and
Dipartimento di Fisica e Astronomia “Galileo Galilei”, Univerisità di Padova, Vicolo dell’Osservatorio 3, I-35122, Padova, Italy 
   \and
 INFN–Padova, Via Marzolo 8, I–35131, Padova, Italy
 \and 
 Department of Astronomy and Theoretical Physics, Lund Observatory, Box 43, SE–221 00, Lund, Sweden
 \and
Max Planck Institute for Astronomy, K\"{o}nigstuhl 17, D69117, Heidelberg, Germany
\and 
Istituto Nazionale di Astrofisica - Osservatorio Astronomico di Padova, Vicolo dell’Osservatorio 5,  I-35122, Padova, Italy  
}

%   \date{Received September 15, 1996; accepted March 16, 1997}

% \abstract{}{}{}{}{} 
% 5 {} token are mandatory
 
  \abstract
   {The degree of mass loss, i.e. the fraction of stars lost by globular clusters, and specifically by their different populations, is still poorly understood. Many scenarios of the formation of multiple stellar populations, %To solve the \enquote{mass budget} problem, many scenarios,
especially the ones involving self-enrichment, assume that the first generation (FG) was more massive at birth than now to reproduce the current mass of the second generation (SG). This assumption implies that, during their long-term evolution, clusters lose around $90\%$ of the FG.
We have tested whether such strong mass loss could take place in a massive globular cluster orbiting the Milky Way at $4 \kpc$ from the centre and composed of two generations. We perform a series of $N$-body simulations for $12 \Gyr$ to probe the parameter space of internal cluster properties. We have derived that, for an extended FG and a low-mass second one,
the cluster loses almost $98\%$ of its initial FG mass and the cluster mass can be as much as 20 times lower after a Hubble time. Furthermore, under these conditions, the derived fraction of SG stars, $f_{\rm enriched}$, falls in the range occupied by observed clusters of similar mass ($\sim 0.6-0.8$).
In general, the parameters that affect the most the degree of mass loss are the presence or not of primordial segregation, the depth of the central potential, $W_{0,FG}$, the initial mass of the SG, $M^{ini}_{SG}$, and the initial half-mass radius of the SG, $r_{h,SG}$. Higher $M^{ini}_{SG}$ have not been found to imply higher final $f_{\rm enriched}$ due to the deeper cluster potential well which slows down mass loss.
}

   {}
   {}
   {}
   {}

   \keywords{methods: numerical  - globular cluster: general - stars: kinematics and dynamics - Galaxy: evolution  - Galaxy: kinematics and dynamics - Galaxy: disc}

   \maketitle
%
%-------------------------------------------------------------------

\section{Introduction}
\label{sec:intro}

In the last decades, an increasing number of observations have revealed the presence of multiple stellar populations (MPs) within globular clusters (GCs), a discovery that has revolutionized our view of these stellar systems \citep{gratton2019}. Stars belonging to distinct populations differ in their light element abundances (such as C, N, O, Na, Mg and Al) while they share, at least in the bulk of GCs, the same iron content. These variations are well defined and linked by anticorrelations like the C-N, Na-O and Mg-Al ones \citep{piotto2005,carretta2009a,milone2017,gratton2019,masseron2019,marino2019}. In particular, within the same GC we can distinguish between stars sharing the same chemical composition of the field ones (O-rich and Na-poor), labelled as first population, and O-poor and Na-rich stars classified as second population.
However, different populations do not differ only in their chemical abundances but also in their structural and kinematical properties, suggesting a deep connection between the origin of the chemical imprints in MPs and the formation and the subsequent dynamical evolution of the whole cluster.

Although the long-term dynamical evolution these systems have undergone is gradually erasing the structural and kinematical differences that MPs had at birth, dynamically young clusters may retain some memory of the original differences between MPs until the present day, in particular in their outskirts \citep{vesperini2013}. By means of observational data analysis, supported by $N$-body models, \citet{dalessandro2019} have shown the tight connection between the relative degree of concentration of different populations and the evolutionary stage of the cluster. In particular, larger differences in the spatial radial distributions between distinct populations are found in clusters that are dynamically young and have experienced lower mass loss. 
The difference between the two populations is not restricted to the spatial distribution: observational studies have revealed that second-population stars are, in some clusters, characterized by a more radially anisotropic velocity distribution \citep{richer2013,bellini2015,bellini2018,libralato2019,libralato2023}, a more rapid rotation \citep{lee2015,lee2017,cordero2017,lee2018,dalessandro2019,kamann2020,cordoni2020,szigeti2021}, a lower fraction of binaries \citep{dorazi2010,lucatello2015,milone2020b} and are more centrally concentrated \citep{norris1979,sollima2007,lardo2011,milone2012b,richer2013,cordero2014,simioni2016,dondoglio2021} than the first population.

In addition to the different degrees of variations in the structural and kinematical properties between MPs, more massive clusters are generally found to host a higher fraction of second-population stars (up to 90\%) than lower mass ones (down to 30-40\%, even 10\% in the Magellanic Clouds, see e.g. \citealt{milone2022}) \citep{milone2017,zennaro2019,dondoglio2021}. 
This quantity is the result of a complex combination of formation history and evolutionary effects since due to the differences between the two populations at birth, they will experience distinct dynamical evolution and, therefore, distinct mass loss rates, which will imply a change of the fraction of second population with time.

Despite a large amount of observational and theoretical studies providing new insights on the chemical and kinematical properties of MPs, a clear understanding of how globular clusters were formed is still not reached \citep{renzini2015,bastian2018,gratton2019}. One of the crucial points deals with the origin of the processed material and the consequent formation of \enquote{anomalous} stars out of it. Many scenarios have been suggested in order to tackle this issue proposing different sources of the processed gas such as asymptotic giant branch stars (AGB) \citep{dercole2008,dercole2016,bekki2017, calura2019}, fast-rotating massive stars \citep{decressin2007}, massive stars \citep{elmegreen2017}, supermassive stars \citep{denissenkov&hartwick2014,gieles2018}, massive interacting binaries \citep{demink2009,bastian2013,renzini2022}, black holes accretion discs \citep{breen2018} and stellar mergers \citep{wang2020}. Nevertheless, so far, none of these scenarios is able to reproduce all the available observational constraints and therefore demand further and thorough developments \citep{renzini2015, bastian2018}.

The physical processes modulating the mass loss in stellar clusters are manifold. Firstly, two-body relaxation was found to gradually set up a Maxwellian velocity distribution, which leads loosely bound stars to overcome the cluster escape velocity \citep{ambartsumian1938,spitzer1940}. Later, \citet{chernoff1990} followed the evolution of multi-mass clusters with a tidal cut-off, driven by two-body relaxation and stellar evolution mass loss. They found that the combination of these two processes leads to a stronger mass loss than the sum of the two independent contributions. The dynamical evolution of a stellar cluster is, however, affected by many other factors, such as binarity \citep{tanikawa2009,fujii2011}, tidal fields \citep{baumgardt2003}, gravitational and tidal shocks \citep{gnedin1997,vesperini1997}, mass segregation \citep{baumgardt2008,vesperini2009,haghi2014} and the presence of dark remnants \citep{contenta2015,banerjee2011,giersz2019}.
All these quantities are, however, known for several present-day clusters, but not for star-forming clusters, posing challenges in setting the initial conditions for simulated clusters. Such uncertainty on the initial values affects also many other parameters, such as the initial mass of the cluster (and also its radial distribution), which would be vital to understand how clusters form and dynamically evolve. Indirect derivations can be obtained starting from clusters' present-day mass and fraction of enriched stars. Assuming that different populations are also distinct generations, and therefore that GCs have undergone self-enrichment, it is possible to define a first generation (FG) composed of normal stars, and a second generation (SG) whose stars possess the peculiar chemical composition. If the mass of the FG is assumed to be comparable to the present-day mass of GCs, one ends up with a mass released by the FG polluters that is much lower than the mass of SG stars observed today, which leads to the so-called \enquote{mass budget problem}. To overcome this problem, it is generally assumed that the cluster, and therefore the FG, was much more massive, between 5 and 20 \citep{decressin2007,dercole2008,schaerer2011,cabreraziri2015}, at its birth but then, during the evolution, most of the FG stars (up to $\sim 95\%$) were lost, so the observed relative number of SG and FG is still reproduced. 

Few attempts have been carried out in order to determine whether, during the long-term evolution, clusters are able to lose a significant fraction of FG stars and then reproduce, after a Hubble time, the observed clusters' features.
The pioneering work on the topic was done by \citet{dercole2008}, who performed a series of $N$-body simulations in the AGB framework, concluding that a cluster with a more concentrated SG generation loses a substantial number of FG stars, at variance with the SG ones, deriving fractions of main sequence stars (MS) $f_{MS}=N_{SG,MS}/N_{FG,MS}$ in agreement with observations.

Later, \citet{bastian2015} %found no correlation between the present-day fraction of SG, which lies between $50$ to $90\%$, and various cluster parameters such as the Galactocentric distance, the metallicity, and the cluster mass. They also 
showed that, combining the observational data with the results of the $N$-body studies of \citet{baumgardt2003} and \citet{khalaj2015}, no match was found, concluding that neither gas expulsion nor the effect of tidal fields could lead to the present-day fraction of SG.

By means of Monte Carlo simulations, \citet{vesperini2021} and \citet{sollima2021} have studied the dynamical evolution of a cluster composed of two populations and a mass of $\sim 10^6 {\rm M_{\odot}}$. Similarly to \citet{dercole2008}, they find that the cluster loses more FG stars and reaches, after $13 \Gyr$, the typical values of SG fraction observed in present-day GCs.
Similar results were also obtained by \citet{sollima2022}, who focused on the binary fractions of the populations. They concluded that the present-day SG binary fraction can be used to constrain the initial concentration of SG stars, providing a relation between the initial size of the SG and total cluster mass.

From E-MOSAICS cosmological simulations, \citet{reinacampos2018} explored the impact of dynamical cluster disruption of multiple stellar populations deriving the degree of mass loss and the fraction of enriched stars as a function of cluster mass, Galactocentric distance, and metallicity. They found discrepancies with observations and therefore concluded that mass loss is unlikely to have a strong impact on shaping the present-day GCs. They also derived that, to reconcile the observations, a significantly larger half-mass radius has to be assumed at birth, and a higher initial SG fraction than the currently adopted ones would be necessary.

Although the fraction of enriched stars is a very strong constraint widely used to compare simulated clusters with observed ones, other fundamental pieces of information can be extracted from the unbound stars \citep{arunima2023}. \citet{larsen2012} found that around 1/5 of the metal-poor stars in the Fornax dwarf spheroidal galaxy belong to the four GCs, meaning that these GCs could have been, at most, 5 times more massive at their birth, posing a strong upper limit on the fraction of stars that could have been lost by GCs. Besides Fornax GCs have been found to resemble the Galactic ones \citep{larsen2014}, and therefore they could have shared a common origin and evolution, stars initially belonging to Fornax GCs could have been lost in the early phases, therefore, loosening the constraints about the degree of mass loss suffered by the Fornax GCs \citep{khalaj2016}. %Similar results were, in fact, obtained by \citet{webb2015} who applied the mass function - initial mass relation found through $N$-body simulations to 33 Galactic GCs (see also \citealt{baumgardt2017}).
Based on the stellar chemical composition, several studies have been carried out aimed at determining the contribution that GCs could have given to the formation of the Galactic halo \citep{carretta2010c,martell2010,martell2011,ramirez2012,martell2016}. Recently, \citet{koch2019} have analysed the spectra of halo field giant stars. They found that 2\% of the stars in the sample show the \enquote{anomalous} chemical composition typical of SG stars, in agreement with the previous investigations. In addition, they derived that $11\%$ of the stars in the Galactic halo were formed in GCs. This quantity is however strongly affected by the adopted mass loss rate in the early phases and the number of completely dissolved clusters, reaching up to $40-50\%$ when assuming a mass loss factor, i.e. the ratio between initial and final cluster mass, greater than $10$ \citep{vesperini2010}. Both the fraction of field SG stars and field GC stars are extremely precious, providing further constraints to the models, not only on cluster scales, but also at larger ones, to understand how the Galaxy assembly proceeded.% In a cosmological contest, by means of hydrodynamic simulations, various works have studied the formation and evolution of GCs in a Milky Way-like galaxy over cosmic time. While in \citet[see also \citealt{renaud2017}]{pfeffer2018} the efficiency of cluster disruption is not enough to reproduce the observed GC mass function of the Milky Way, in \citet{li2019} too many clusters are disrupted at $z=0$. One of the major discrepancies, pointed out also by the recent work of \citet{rodriguez2022}, concerns the peak of the mass function, which is significantly below the observed one, meaning that current simulations are either overproducing low mass clusters or inefficiently disrupting them.  

\begin{comment}

mass loss causes:
-tidal fields
-two-body relaxation
-stellar evolution 
-binary 
-form of the orbit
-shape of the Galactic potential
-crossing time
-natal kicks of neutron stars
-gas expulsion
-segregation
\end{comment}

%preferential loss of low mass stars \citet{baumgardt2003}. Correction? look at eq. 14)

In this paper, we aim to derive the degree of mass loss in the two different stellar components, to determine whether there are combinations of initial parameter values that can lead to a significant mass loss, as the one required to solve the mass budget problem, and that spawn final clusters compatible with the observed GCs.
We perform a series of direct $N$-body simulations to follow the long-term evolution of a globular cluster with an initial mass of $M\sim 10^7 {\rm M_{\odot}}$ and composed of two populations taking into account stellar evolution, the tidal effects of the Galactic potential, and primordial segregation. Although GC mass loss has been explored in the past, only few works modelled a GC composed of more than one population (i.e. \citealt{vesperini2021,sollima2021,sollima2022}, with Monte Carlo codes). Our simulations are among the first of this kind performed with a direct $N$-body code in the literature, together with the ones of \citet{dercole2008} and \citet{henaultbrunet2015}, where, however, a cluster of lower mass has been considered. The cluster is composed of two stellar populations, and it is assumed to orbit the Milky Way (MW). From the results derived by \citet{calura2019}, confirmed also by \citet{lacchin2022}, the fraction between FG and SG is larger than the present-day ones as assumed to solve the mass budget problem. Here, we aim at quantifying the mass loss factors that are needed to reproduce the observed SG fraction for such a massive GC.
The cluster is located in the disc of the Milky Way and, therefore, is meant to represent a cluster belonging to the in-situ population.
There are various reasons why studying disc GCs is important. First, the distribution of metal-rich MW GCs are more concentrated and flatter than the metal-poor component, and generally, they are associated with the thick disc and bulge populations \citep{armandroff1988,armandroff1989,zinn1985,minniti1995,cote1999,vandenbergh2003,bica2006,bica2016}. In addition, disc GCs, which now constitute almost one-third of the total MW GCs \citep{harris2010}, could have been much more in the past. Field stars showing GC-like features have been discovered in the inner Galaxy \citep{schiavon2017,fernandeztrincado2022}, a detection supported by simulations showing that tidal effects in the inner regions of MW-like galaxies could have gradually destroyed disc GCs, decreasing their population \citep{renaud2017}. Lastly, kinematic heating due to several accretion events could have also deprived the disc of GCs, which would now be part of the inner halo \citep{kruijssen2015,dimatteo2020}.

The paper is organized as follows: in Section \ref{sec:methodsnb}, we describe the model we are adopting and the novelty introduced in the present work. Section \ref{sec:resultsnb} deals with the results we have obtained for our sets of simulations. In Section \ref{sec:discussionnb}, we discuss the outcomes of the simulations and compare them with the literature and observations. Finally, we draw our conclusions in Section \ref{sec:conclusionsnb}.

%altri articoli sulla mass loss in multiple stellar pop con N-body o monte carlo \citep{sollima2022}

\begin{figure*}
        \centering

        \includegraphics[width=0.9\textwidth,trim={0.2cm 0.2cm 0.2cm 3cm},clip]{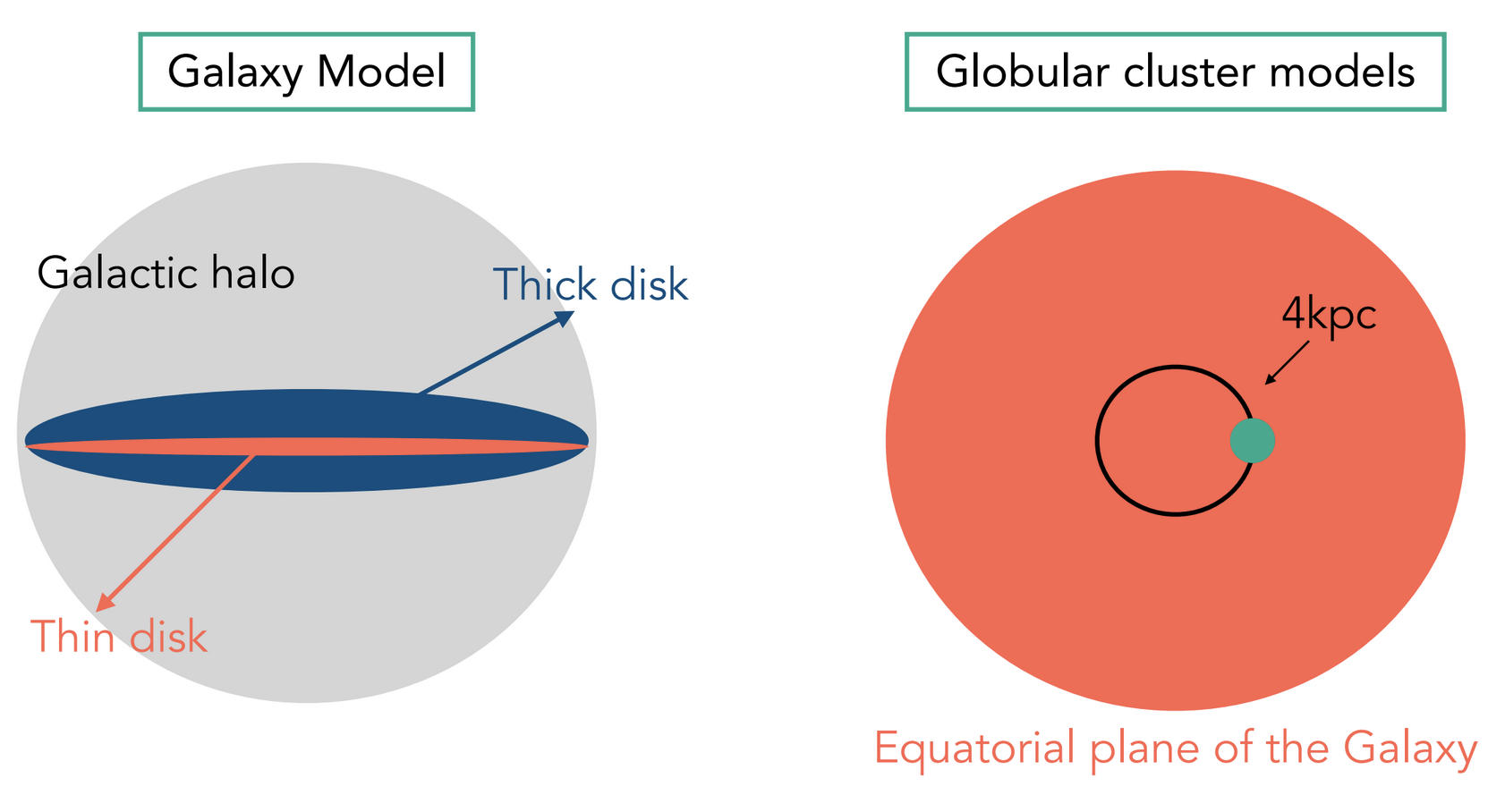}  
        
        \caption{On the left, a schematic representation of the Galaxy model adopted for the simulations, which includes both the thick and thin disc plus the halo, while the bulge is modelled as part of the Galactic disc. The right panel illustrates the GC model. The simulated cluster is located at $4 \kpc$ from the Galactic centre and is assumed to orbit it on the plane of the disc. Credits: Alessandra Mastrobuono-Battisti.}
  \label{fig:model}
\end{figure*}

\section{Models and Method}
\label{sec:methodsnb}

We study the internal evolution of a series of cluster models, including either one or two stellar populations, firstly without star formation and stellar evolution and then including these ingredients to study their effect on mass loss. In the next subsections, we provide details on the initial setups adopted for our simulations, illustrating the assumptions we adopted depending on the characteristics of the model and the phenomena that we explored.

\subsection{Description of the code }
We ran our simulations using an updated version of NBSymple \citep{capuzzodolcetta2011}\footnote{See \href{https://github.com/alessandramb/NBSymple}{https://github.com/alessandramb/NBSymple} for the basic version of the code}, a direct and symplectic $N$-body code parallelized on GPUs. Several versions of this code are available and have been used to study various aspects of GCs evolution in the Galactic potential (see e.g. \citealt{mastrobuonobattisti2012,sollima2012,leigh2014,mastrobuonobattisti2019}). This new version includes a star formation and stellar evolution routine for the first and second generations separately. 
The softening length adopted to avoid close encounters is equal to the mean interparticle distance within the half-mass radius as done by \citet{dercole2008} and it is calculated separately for the first and second generations. Our choices for the softening length and the number of particles used to represent the clusters are motivated by the computational limitations, as we aimed at 12 Gyr long simulations. 
We have tested the dependence of our results on the choice of the softening length by performing a simulation with a softening length 10 times lower than the mean interparticle distance, with the timestep modified accordingly as in \citet{mastrobuonobattisti2012}. We have obtained that the cluster properties are only very weakly affected, in particular the fraction of SG stars, $f_{\rm enriched}$. Our simulations are not meant to model close
encounters and hard binaries: to account for these effects we would need a substantially higher number of particles, smaller softening length and shorter timestep, which would make  12 Gyr simulations computationally unfeasible. However, we recall that our simulations start when first-generation massive stars have already exploded. In addition, in order to avoid iron pollution, second-generation stars are assumed to be
composed of only low and intermediate-mass stars. Therefore, massive interacting binaries
do not take part in our models either because they have already evolved or they never form.
Lower-mass binaries, neglected in our model, but expected to be present in real systems, would increase the number of ejected stars \citep{kupper2008}. However, while binaries heat up the system, a smaller softening makes the cluster more compact, limiting its ability to lose stars. In any case, all these collisional effects are expected to influence the mass loss much less than those related to stellar evolution, which are accounted for in our models.
%, whose effects would be significant with softening lengths $10^4$ times smaller than the interparticle distance (assuming a scale for tight binaries of the order of 100 AU). In our symplectic code, the timestep is fixed and equal for all the particles, and is proportional to the softening length $\epsilon$ as $\Delta t= \sqrt{ \epsilon^3/(GM)}$, which would be of the order of $10^{-2} {\rm yr }$ and, therefore, computationally unfeasible. With a softening length smaller than ours, collisions would be more effective in driving mass loss; however, the smaller the softening length, the more compact the cluster becomes, limiting its ability to lose stars. The intensity of mass loss would therefore depend on which of these two processes prevails, which is hard to predict and would be an essential future step to be done.}

The mass lost due to stellar evolution is instantaneously removed and, therefore, energy and angular momentum are not conserved.

\subsubsection{Galactic potential model}
Our Galactic model consists of a dark matter halo with both a thin and thick disc, as shown in the left panel of Figure \ref{fig:model} (see also \citealt{mastrobuonobattisti2019}). The functional forms of these components are taken from \citet{allen1991} with the parameters from Model II of \citet{pouliasis2017}, which aims at reproducing the actual MW. Such a model is able to reproduce various observables, such as the rotation curve, thin and thick disc scale lengths and heights, and stellar density in the solar neighbourhood.
The bulge is considered as part of the Galactic disc, so it is not represented as an independent component \citep{dimatteo2016}. The mass assumed for the halo is $2.07\times 10^{11} {\rm M_{\odot}}$ with a scale height of $ 14{\rm \ kpc}$. The thick disc has a mass of $3.91\times 10^{10} {\rm M_{\odot}}$ with radial and vertical scale length of $2{\rm \ kpc}$ and $800 {\rm\  pc}$, while the thin disc has a mass of $3.68\times 10^{10} {\rm M_{\odot}}$, a radial scale length of $4.8 {\rm \ kpc}$ and a scale height of $250 {\rm \ pc}$.  With the adopted analytic Galactic model, we do not
take into account the dynamical friction, which would be self-consistently included if the MW had
been modelled as an $N$-body system, thus composed of stellar and dark matter
particles. To assess the importance of this neglected process, we have estimated, through Equation 8.13
of \citet{binney2008}, that for a satellite of mass $8.6\times 10^6\Msun$ the timescale of dynamical friction at $4 \kpc$ from the Galactic centre is $t_{\rm df}=80\Gyr/{\rm ln}\Lambda$ where ${\rm ln}\Lambda $ is the Coulomb logarithm, assuming $M(<4\kpc)=4\times 10^{10}\Msun$ for the mass of the MW within $4\kpc$ derived through the model of \citet{pouliasis2017}. Despite the spherical approximation \footnote{see \citealt{bonetti2021} for dynamical friction calculations for disc structures}, this result suggests that dynamical friction is negligible for the systems we are modelling.

\subsection{Single stellar population models}
Our single population clusters are modelled using the \citet{king66} profile, adopting different values for the half-mass radius, $r_h$, and for the dimensionless central potential $W_0$ \citep{binney2008}, which varies between $2$ and $7$. In this way, we explore the behaviour of both loose and dense clusters. The cluster initial mass is, for all models, ${\rm 10^7\,M_\odot}$ , with an initial metallicity of Z=0.001, since, as explained before, we aim at modelling very massive clusters, like the ones in \citet{calura2019} and \citet{lacchin2021,lacchin2022}. The initial metallicity of the cluster is  Z=0.001 as in \citet{calura2019}. 
The initial positions and velocities of the particles, in the absence of the external gravitational galactic potential, are derived using the software NEMO, through the $mkking$ routine \citep{teuben1995}.
Our clusters are single-mass models (i.e. all stellar particles have the same mass) and are represented either with $N=102400$ or $N=25600$ particles, meaning that each particle is significantly more massive than a star. This choice is due to current computational limitations in running direct $N$-body simulations of systems with $10^6$ or more particles for a timespan of 12 Gyr. 
%{\bf Both the undersampling and the adoption of single-mass stars have implications on the mass loss suffered by the clusters. The undersampling leads to a lower relaxation time, which scales with the number of particles. This would possibly lead to a stronger mass loss; however, as we will see later, the major driver of mass loss is stellar evolution rather than relaxation. On the other hand, by adopting single-mass particles we are not modelling mass segregation from binaries and massive star remnants which would sink into the innermost regions of the cluster. Therefore, adopting a mass spectrum rather than single-mass particles would mostly affect the center of the system and possibly increase mass loss \citep{kupper2008}.} 

%Both the undersampling and 
The adoption of single-mass stars has implications on the mass loss suffered by the clusters. %The undersampling leads to a lower relaxation time, which scales with the number of particles. 
We are not modelling mass segregation from binaries and massive stars remnants, which would sink into the innermost regions of the cluster, eventually favouring the loss of low-mass stars. However, the major driver of mass loss is stellar evolution, which also induces mass loss due to the shallowing of the potential, so we do not expect that the main conclusions of this work would be significantly different if one considered simulations accounting for the presence of a mass spectrum.% and with a larger number of particles. 

Each cluster in our simulation starts with an actual mass that is slightly lower than ${\rm 10^7\,M_\odot}$. Since the simulation starting point is set after the explosion of FG core-collapse supernovae (i.e. at a time equal to $t_0= 30$\,Myr), we remove $16\%$ of the initial cluster mass reaching $M_0=8.4\times 10^6 {\rm M_\odot}$, to mimic the effect of the death of massive stars. This value is the mass return fraction due to the evolution of stars with a mass larger than ${\rm 8\,M_\odot}$ for the \citet{kroupa2001} initial mass function (IMF) with a final-to-initial mass relation from \citet{agrawal2020}.

Due to this mass removal, the system goes out of virial equilibrium and will expand to return to an equilibrium state. %Such mild expansion can be interpreted as the remnant of the violent relaxation already suffered by the cluster. 
In our best model, this relaxation leads to a half-mass radius increase of 7\% and no significant change in the mass loss, since the relaxation is taking place in the inner region, while the outskirts are very weakly affected.

\begin{table*}
\caption{Models with neither stellar evolution nor SG stars run in this work. The initial mass of each simulated cluster is $M_0={\rm 8.4\times 10^6 M_\odot}$, after the removal of $16\%$ of its mass. }             % title of Table
\label{table:1}      % is used to refer this table in the text
\centering                          % used for centering table
\begin{tabular}{c c c c c  c c}        % centered columns (4 columns)
\hline\hline                 % inserts double horizontal lines
Model$^{(a)}$  & $W_0$ & $r_{h}$& Segregation  & Number of particles & Softening length & $f_{mass\ loss}$ \\    % table heading 
      &      &  (pc)   &  &   &(pc)          \\    % table heading 
\hline                        % inserts single horizontal line

   %%% I valori della perdita di massa sono in parte errati e mancano alcuni modelli!!! Apetta a trarre conclusioni.
   n7N & 7    &23& N  & 25600   & 1.65 &  0.44 \\
   n7Y  & 7    &23& Y  & 25600   & 1.65 & 0.58  \\
   n5N  & 5    &37& N  & 25600   & 2.57 & 0.50\\
   n5Y  &  5   & 37&N  & 25600   & 2.57 &  0.57 \\
   n2N  & 2    &60& N  & 25600   & 4.20 & 0.72 \\
   n2Y  & 2   & 60&Y  & 25600 & 4.20 &  0.79 \\
   N2N  & 2    & 60& N  & 102\,400  & 2.65 &  0.57\\ %running

\hline                                   %inserts single line
\end{tabular}
\label{tab:FGonly_massloss}
\\
\raggedright $^{(a)}$Model name: n/N = small/large number of particles +  $W_0$ + N/Y = non-segregated/segregated.
\\

 {\it Columns:} 1) Name of the model; 2) the adimensional central potential parameter $W_0$ of the FG; 3) half-mass radius of the FG; 4) primordial segregation of the FG (N = not segregated, Y = segregated), 5) the number of particles $N_{tot}$, 6) softening length, 7) mass loss fraction defined as $f_{mass\ loss}=(M_{ini}-M_{fin})/M_{ini}$.
\end{table*}

%%%%%%%%%%%%%%%%%%%%%%%%%%%%%%%%%%%%%%%%%%%%%%%%%%%%%%%%%%%%%%%%%%%%%%%%%%%%%%%%%%%
%%%%%%%%%%%%%%%%%%%%%%%%%%%%%%%%%%%%%%%%%%%%%%%%%%%%%%%%%%%%%%%%%%%%%%%%%%%%%%%%%%%
%%%%%%%%%%%%%%%%%%%%%%%%%%%%%%%%%%%%%%%%%%%%%%%%%%%%%%%%%%%%%%%%%%%%%%%%%%%%%%%%%%%

\begin{table*}
\caption{Models with simplified stellar evolution and without SG stars run in this work. The initial mass of each simulated cluster is $M_0={\rm 8.4\times 10^6 M_\odot}$, after the removal of $16\%$ of its mass. }             % title of Table
\label{table:1}      % is used to refer this table in the text
\centering                          % used for centering table
\begin{tabular}{c c c c c c c}        % centered columns (4 columns)

\hline\hline                 % inserts double horizontal lines
Model$^{(a)}$ &  $W_{0}$& $r_{h}$ & Segregation  & Number of particles & Softening length& $f_{mass\ loss}$ \\    % table heading 
      &       & (pc)    &  & & (pc)           \\    % table heading 
\hline                        % inserts single horizontal line

   %%% I valori della perdita di massa sono in parte errati e mancano alcuni modelli!!! Apetta a trarre conclusioni.

   n7Ne &  7   & 23 & N  & 25600   & 1.65 & 0.63  \\
   n7Ye &  7    & 23& Y  & 25600  & 1.65 & 0.74 \\
   n5Ne & 5     &37 & N  & 25600   & 2.57  & 0.70 \\
   n5Ye &  5    &37& Y  & 25600   & 2.57  & 0.79  \\
   n2Ne  & 2 &60   & N  & 25600   & 4.20 &  1.00\\ %%missing!
   n2Ye & 2&60    & Y  & 25600   & 4.20 &  1.00  \\
   N2Ne &  2 &60   & N  & 102\,400  & 2.65  & 1.00 \\ 
   
\hline                                   %inserts single line
\end{tabular}
\label{tab:stev_massloss}
\\
\raggedright $^{(a)}$ Model name: Model name: n/N = small/large number of particles +  $W_0$ + N/Y =  non-segregated/segregated + e = with stellar evolution.
\\
 {\it Columns:} same as in Table \ref{tab:FGonly_massloss}.% 1) Name of the model; 2) $W_0$ of the FG; 3) half-mass radius of the FG; 4) primordial segregation of the FG (N = not segregated, Y = segregated), 5) number of particles $N_{tot}$, 6) softening, 7) mass loss fraction defined as $f_{mass\ loss}=(M_{ini}-M_{fin})/M_{ini}$.

\end{table*}

%% in realta' per questi modelli avevo usato il 13% ma non credo faccia molta differenza
We explore both clusters with and without primordial mass segregation.

Primordial mass segregation has been found to have a significant effect on the cluster mass loss due to the cluster expansion in response to the massive star mass loss, happening preferentially at the cluster centre \citep{vesperini2009,haghi2014}. 
In case the clusters are mass segregated, we use the software McLuster \citep{kupper2011} to calculate the radius comprising all the massive stars in a primordially segregated model. We then remove the mass that is lost due to the explosion of stars more massive than $8\ {\rm M_\odot}$ within this radius, keeping a King profile for the density. 
All the models orbit the Galaxy in the plane of the disc at a galactocentric distance of $4 {\kpc}$, as shown in the right panel of Figure \ref{fig:model}. 
This is the same distance assumed by \citet{dercole2008}, albeit they used a simpler model, including a point-like mass located at the galaxy centre. It is worth mentioning that in our model, the mass enclosed within a radius of 4 kpc, derived integrating the density distribution, is $4.0\times 10^{10} \Msun$. This is in very good agreement with the mass assumed for the point-mass galaxy potential assumed by \citet{dercole2008}. 
The clusters are tidally filling, i.e. their tidal radius, $r_t$, is equal to the distance at which the cluster potential and the Galactic potential have the same value \citep{vonhoerner1957,baumgardt2003,webb2013}. %%aggiungere altre referenze
As the tidal radius is fixed to ${ 200 \pc}$, the core radius and half-mass radius of each of the models vary depending on the value of the $W_0$ parameter.

\subsubsection{Without long-term stellar evolution}

We initially modelled clusters hosting a single stellar population, that are only affected by dynamical effects (i.e. not considering any long-term stellar evolution effect). To start with the same cluster mass, we here remove the 16\% of the initial mass, as we will do in all the other models. The details on the models are reported in Table \ref{tab:FGonly_massloss}.

%%\textcolor{magenta}{Mancano i modelli con solo FG e stellar evolution.}
\subsubsection{Adding long-term stellar evolution}
In our second set of models, we still have only one stellar population, but we considered the effects of long-term stellar evolution. This is done through a mass return fraction taken from the relation between remnant mass and progenitor of \citet[case METISSE with MESA of Fig. 7]{agrawal2020} given by:
 \begin{equation}
\begin{split}
    m_{loss}(t)=m_p(t=0)[b_0+b_1 {\rm log}(t)+b_2 {\rm log}^2(t)]
\label{eq:MRFFG}
\end{split}
\end{equation}
where $m_p(t=0)$ is the mass of the particle before the removal of the $16\%$ of the mass due to massive star winds and SN explosions, $b_0=0.329420$, $b_1=0.0379353$, $b_2=-0.002760463$ and $t$ is expressed in ${\rm Gyr}$. At $t= 0.03 {\rm Gyr}$, the mass lost is the $16\%$ of the whole mass, which is the mass we removed to take into account the death of massive stars.
%
%This represents an analytical fit to the stellar mass return of the low and intermediate-mass stars models of \citet{vandenhoek1997}. 

It is worth noting that at metallicity $Z>10^{-3}$, as in our case, the variations in the fractional cumulative mass loss (expressed by the \enquote{returned fraction}) are of the order of a few percent \citep{vincenzo2016}.

The parameters adopted for these models can be found in Table \ref{tab:stev_massloss}.

%%%%%%%%%%%%%%%%%%%%%%%%%%%%%%%%%%%%%%%%%%%%%%%
\subsection{Two stellar populations models}\label{sec:newmod}

In our third set of simulations, we finally add the SG, embedded inside the FG component. All the particles in the system have the same mass, for a total number of particles $N_{tot}=102400$. Both components are spherical and represented by \citet{king66} models. The FG component, modelled with $N_{FG}$ particles, is a King model with $W_{0,FG}$ ranging between $2$ and $7$, a total initial mass of $M^{ini}_{FG}=10^7\ {\rm M_\odot}$ and a tidal radius of $200\ {\rm pc}$, to mimic a tidally filling system. As before, we build the initial positions and velocities using the software NEMO, through the $mkking$ routine \citep{teuben1995}.

%--------------------------------------
   \begin{figure*}
   \centering
      \includegraphics[width=0.45\textwidth,trim={34.cm 0 0 0},clip]{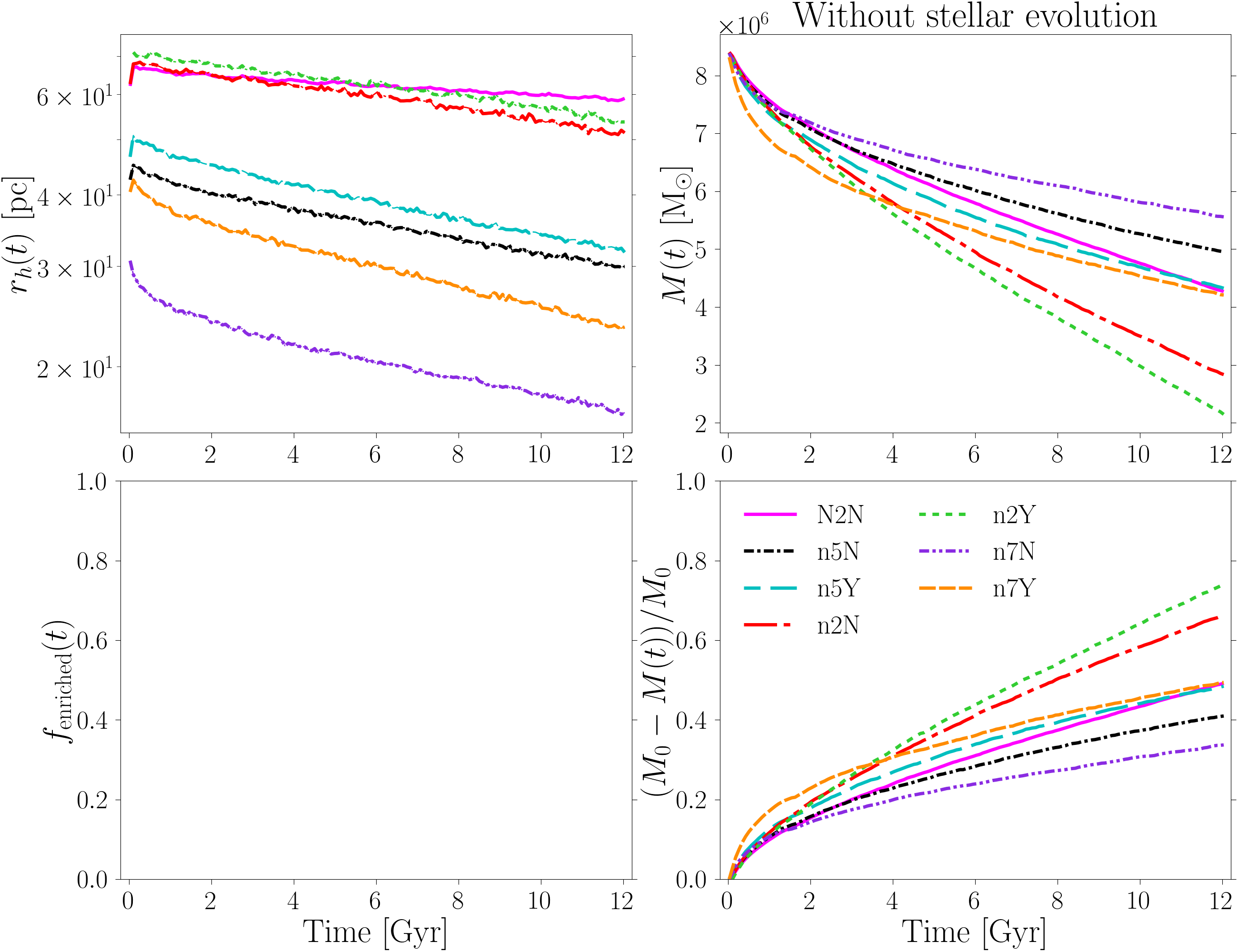}
      \includegraphics[width=0.46\textwidth,trim={ 33.2cm 0 0 0},clip]{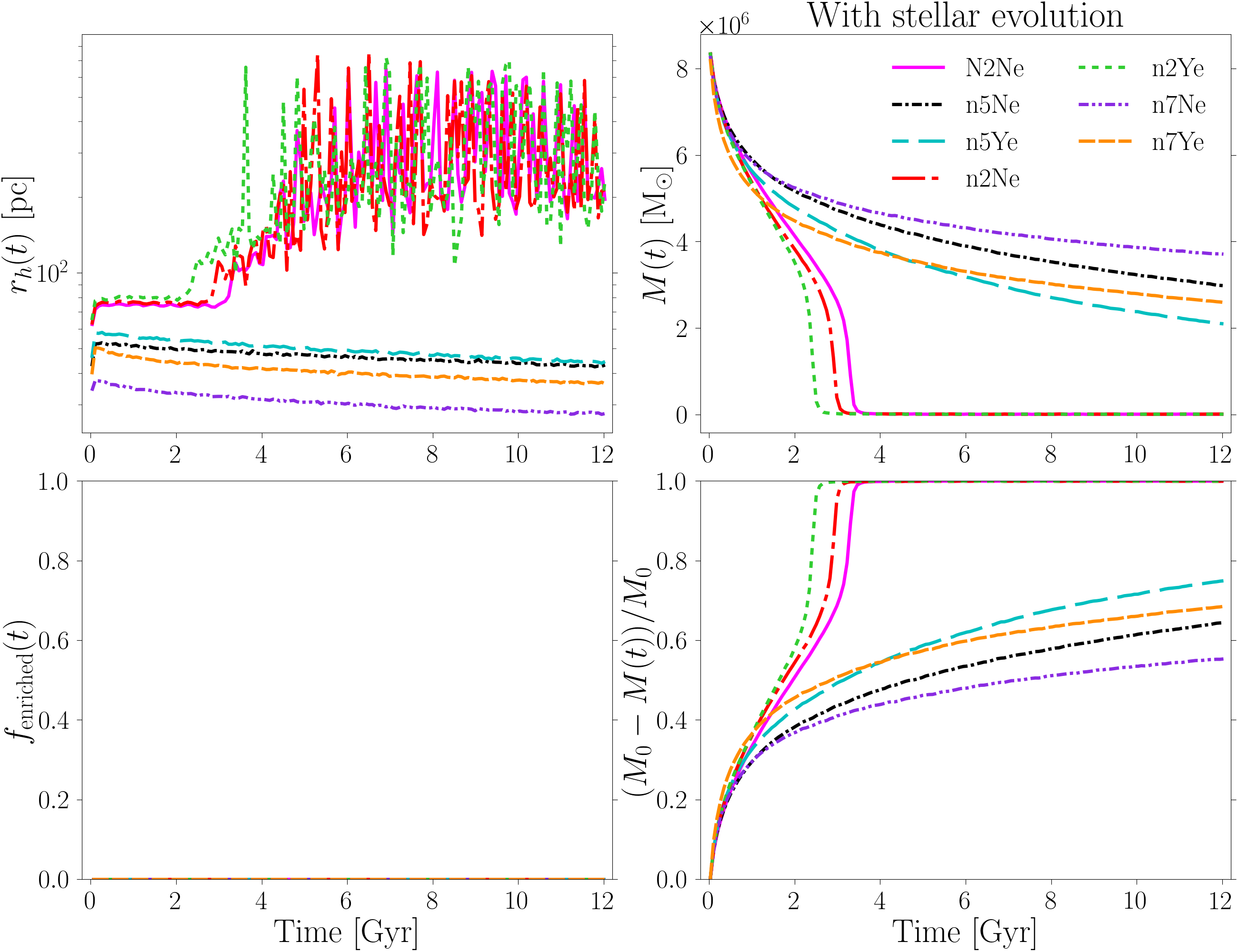}
      
   \caption{Mass (top) and mass loss (bottom) as a function of time. The left column represents the models with the FG only and without stellar evolution, listed in Table \ref{tab:FGonly_massloss}, while, in the right column the models with only FG but where stellar evolution is taken into account, listed in Table \ref{tab:stev_massloss}, are shown.}
              \label{fig:FGonly_massloss}%
    \end{figure*}

The SG component is a King model where we vary $W_{0,SG}$ from $5$ to $7$, its mass from $M^{ini}_{SG}={\rm 7\times 10^5\ M_{\odot}}$ to  ${\rm 3\times10^6\,M_\odot}$ (as a consequence, also the number of SG particles $N_{SG}=N_{tot}-N_{FG}$ will change from $7877$ to $26947$, respectively) and the half-mass radius from $1$ to $6$ pc. We vary the SG mass to test different initial SG fractions. We vary the velocity dispersion of the SG as well, to explore the effect of this parameter on the cluster's mass loss rate. We run models with different values of the central velocity dispersion, equal to $ 0$, ${\rm 10\ km \ s^{-1}}$ and to the velocity dispersion of the generated King model, where, for the first two values, we rescaled the velocities derived for the King. In the third case, the SG is in equilibrium as an isolated system, while, in the other cases, it is radially out of equilibrium and it tends to collapse and readjust after a phase of violent relaxation.

As before, the mass of the FG at the beginning of the simulation is slightly lower than its initial mass, since $16\%$ of the mass is removed due to the explosions of core-collapse supernovae.
After that, the FG starts to evolve dynamically as FG stars lose their mass due to stellar evolution, with a cumulative mass return fraction given by Equation \ref{eq:MRFFG}. 

The SG appears after $10\ {\rm Myr}$ from the beginning of the simulation (i.e. at a time of $t_1= 40\ {\rm Myr}$) and grows its mass at a constant star formation rate of ${\rm 0.05\,M_\odot\ yr^{-1}}$ (for $M^{ini}_{ \rm SG}=3\times 10^6\ { \rm M_{\odot}}$; ${\rm 0.01\,M_\odot\ yr^{-1}}$ for $M^{ini}_{ \rm SG}=7\times10^5\ { \rm M_{\odot}}$) for a total of $60\ {\rm Myr}$ (see \citealt{calura2019}). To avoid the contribution of SG massive stars, which would chemically pollute the AGB ejecta with e.g. iron, we assume, as it is generally done in the AGB scenario, a truncated SG IMF composed only of stars with masses smaller than $8 \ {\rm M_\odot}$ (see \citealt{dercole2010,bekki2019}). SG stars are kept fixed with respect to the cluster centre of density while they are forming. Once the total initial mass of the SG is reached, they start to evolve dynamically. After an additional $30\ {\rm Myr}$, the SG has accumulated enough mass and its stars start evolving following a cumulative mass return fraction law of the same shape of Equation \ref{eq:MRFFG} but rescaled by a factor of 1.27133. The time evaluation of $m_p$ is shifted as well, due to the later formation of the SG, and will correspond to the time at which the SG has stopped growing in mass.
\begin{comment}
{\begin{equation}
\begin{split}
     m_{loss}(t)=1.27133\ m_p(t=0)[b_1 log(t-t_1)+
     b_2 log^2(t-t_1)]
\label{eq:MRFFG2}
\end{split}
\end{equation}

{\begin{equation}
\begin{split}
     m_{loss}(t)=m_p(t)-1.27133 m_p(t)[b_1 log(t-t_1+t_0)+\\
     b_2 log^2(t-t_1+t_0)]+
     1.27133 m_p(t)[b_1 log(t_0)+b_2 log^2(t_0)]
\label{eq:MRFFG2}
\end{split}
\end{equation}}
\end{comment}
%where $t_1$ is expressed in ${\rm Gyr}$ and all the other parameters are as in Equation \ref{eq:MRFFG}.}
The mass is added or removed in equal measure from each star particle in the relevant component of the cluster.

%%%%%%%%%%%%%%%%%%%%%%%%%%%%%%%%%%%%%%%%%%%%%%%%%%%%%%%%%%%%%%%%%%%

\section{Results}
\label{sec:resultsnb}
In this section, we present the results obtained from our simulations. First, we describe the outcomes of models where the FG only is modelled and stellar evolution is not taken into account. Secondly, we report the results of the simulations assuming stellar evolution but still with the FG component only. Lastly, 
the outcomes of the simulations with both stellar evolution, FG, and SG components are described.

%----------------------------------------------------------

\subsection{Models with single stellar population}
\subsubsection{Models without stellar evolution}
We have first studied the long-term evolution of a massive cluster, $M_{ini}=10^7{\rm M_{\odot}}$, composed only by FG stars. At the beginning of the simulation, the stellar mass is equal to $M_0=8.4 \times 10^6 {\rm M_{\odot}}$, which represents the mass of low and intermediate-mass stars plus the remnants of the massive ones left in the system after massive stars have gone off.

\begin{sidewaystable*}
%\begin{table*}
\caption{Models with both simplified stellar evolution and SG stars. The initial FG mass of each simulated cluster is $M_0={\rm 8.4\times 10^6 M_\odot}$, after the removal of $16\%$ of its mass.} %Masses are all in terms of $10^5$.}             % title of Table
\label{table:1}      % is used to refer this table in the text
\hspace{-0.4cm}                          % used for centering table
\resizebox{1.016\columnwidth}{!}{
\renewcommand{\arraystretch}{1.10}
\large
%\begin{tabular}{P{1.9cm}cc P{0.7cm}P{0.9cm}c cP{0.9cm}P{0.9cm} cP{0.75cm}c cc }        % centered columns (15 columns)

\begin{tabular}{ccccc ccccc ccccc}  
\toprule
\multicolumn{1}{c}{} &
\multicolumn{6}{c}{Initial}    &
\multicolumn{8}{c}{Final}    \\ 
\cmidrule(lr){2-7}
\cmidrule(lr){8-15}

Model$^{(a)}$ & $W_{0,FG}$ &$r_{h,FG}$ &Segr  & $M_{SG}$ & $W_{0,SG}$  &$r_{h,SG}$  & $M_{FG}$ & $M_{SG}$ & $f_{\rm enriched}^{r_{h,tot}}\ (f_{\rm enriched})$&$f_{SG}^{\rm lost}$&$r_{h,FG}$&$r_{h,SG}$&$r_{h,tot}$&$log(\rho_c)$\\    % table heading 

      &     &pc  &&\ $ \rm 10^5 M_{\odot}$ &&pc & $ \rm 10^5 M_{\odot}$ &$\rm 10^5 M_{\odot}$& &&pc&pc&pc&${\rm M_{\odot}/pc^3}$          \\    % table heading 

\hline                        % inserts single horizontal line

   %%% I valori della perdita di massa sono in parte errati e mancano alcuni modelli!!! Apetta a trarre conclusioni.
 
   5N7M3.70  & 5 &37    & N  & 30 & 7 & 3.7&   $35.4$ & $18.0$& 0.44 (0.34)&0.16&25.0&12.5&19.9&3.61\\ 
   5N7M3.710& 5   &37 & N   & 30 & 7 &  3.7&  $33.6$  & $18.5$&0.49 (0.35)&0.15&25.5&10.1&18.9&4.08 \\ 
   5N7M3.7K  & 5   &37 & N   & $30$ & 7 & 3.7 & $36.3$  & $19.8$& 0.49 (0.35) &0.14&24.1&9.34&17.3&4.11\\ 
   5Y7M3.7K   & 5  &37 & Y    & $30$ & 7 & 3.7 &   $26.3$  & $19.3$&0.58 (0.42)&0.13&27.4&9.22&18.2&4.41 \\
   5Y7M1K  & 5   &37 & Y    & $30$ & 7 & $1$ &   $22.9$  & $15.7$&0.46 (0.40)&0.16&27.8&22.4&25.4&2.36\\  
   5Y5M1K  & 5   &37 & Y   & $30$ & 5 & $1$ &   $21.0$  & $15.5$&0.48 (0.42)&0.15&27.7&22.4&25.2&2.36 \\ 
   5Y7m1K   & 5   &37 & Y    & $7$ & 7 & $1$&  $21.9$  & $4.42$ &0.23 (0.16)&0.03& 32.1&20.2&29.9&2.15\\ 
   2N7M1K   & 2   &60 & N  & $30$ & 7 & $1$ &  $12.6$  & $14.1$ &0.58 (0.52)&0.15&24.4&19.9&21.8&2.34\\ 
   4N7M4.5K  & 4 &45& N  &$30$ & 7 & $4.5$&  $27.8$  & $18.8$& 0.58 (0.40)&0.13&29.3&7.84&19.4&4.90 \\ 
   2N7m6K  & 2 &60& N&$7$ & 7 & $6$ &   $3.02$  & $3.33$ & 0.56 (0.52)&0.04&13.2&10.8&11.8&2.55 \\ 
   2N5m6K & 2 &60& N&  $7$ & 5 & $6$ &  $2.93$  & $3.50$ &0.59 (0.54)&0.03&12.9&10.2&11.3&2.45 \\ 
   2N7m1K & 2 &60& N &$7$ & 7 & $1$ &   $ 1.89$  & $2.00$ & 0.59 (0.52)&0.05&10.9&6.19&8.32 &4.98\\ 
   2N5m1K & 2 &60&N &$7$ & 5 & $1$ &   $2.15$  & $2.09$ & 0.57 (0.49)&0.05&10.7&6.11&8.25&4.47 \\ 

\toprule                                  %inserts single line
\end{tabular}\label{tab:SG_massloss} }
\\
\raggedright $^{(a)}$ Model name: $W_{0,FG}$ + Y/N = with/without segregation + $W_{0,SG}$ + m/M = low/high initial $M_{SG}$ + initial $r_{h,SG}$ + $\sigma_{0,SG}$ (K=King velocity dispersion) 
\\
 {\it Columns:} 1) Name of the model; 2) $W_0$ of the FG; 3) half-mass radius of the FG; 4) Primordial segregation of the FG (N = not segregated, Y = segregated), 5) mass of the SG; 6) $W_0$ of the SG; 7) half-mass radius of the SG; 8) the total mass of FG bound stars; 9) the total mass of SG bound stars; 10) fraction of SG stars within the half-mass radius of the whole cluster (fraction of SG stars of the whole cluster); 11) fraction of SG stars among unbound stars $f^{lost}{SG}=M^{lost}{SG}/M^{lost}{tot}$, where $M^{lost}{SG}$ and $M^{lost}_{tot}$ are the mass of unbound SG and the total mass of unbound stars, respectively; 12) half-mass radius of the bound FG; 13) half-mass radius of the bound SG; 14) half-mass radius of the whole cluster (i.e. only bound stars); 15) central density.
 %}
%\end{table*}
\end{sidewaystable*}

In Table \ref{tab:FGonly_massloss}, we summarise the main parameters of our models together with the resulting mass loss fraction $f_{mass\ loss}=(M_{ini}-M_{fin})/M_{ini}$, with $M_{fin}$ the final mass of the whole cluster, at the end of each simulation. Stellar evolution is not taken into account for the moment. Figure \ref{fig:FGonly_massloss} shows the evolution of the cluster mass, on the top left, and of the normalized mass loss, at the bottom left.
As one can expect, the shallower the potential well, and therefore the lower the values of $W_0$, the greater the mass loss at the end of the simulation. However, model $n7Y$, which is characterized by $W_0=7$, is losing more mass than models with lower $W_0$ values, in the first few Myr. 
The higher concentration coupled with initial segregation is responsible for this behaviour, which affects also other models described throughout the section. %\textcolor{red}{non mi e molto chiaro perche si espande di piu. l'effetto della segregazione e' piu' forte quando il cluster e' pi' compatto? lo si vede anche in modelli con SG}
Initial segregation is, in general, leading to a stronger mass loss since segregated systems will have, after the death of massive stars, a larger $r_h$, making the system less bound. 
Apart from varying physical parameters, we have also changed the number of particles from 25600 to 102400 retrieving that, when more particles are used, as in $N2N$, the cluster loses less mass as a result of %both the deepening of the potential well due to the smaller softening length and 
the longer relaxation time.%, which scales as $N/log(N)$, with $N$ the number of particles.
%Model 2YP26 shows the dramatic end of a cluster with a shallow potential well combined with initial segregation. The cluster loses all its mass dissolving after 9 Gyr and dispersing its stars along its orbit. Once the softening length is changed, variations in the mass loss are visible only for large values of this parameter. In particular, the larger the softening length, the more mass is lost by the cluster. Nevertheless, models 7Np16 and 7Np08, although having a different mass evolution, end up with very similar mass after 12 Gyr.
%\textcolor{magenta}{Dire che abbiamo variato epsilon per verificarne l'effetto.}

%-----------------------------------------------------------

%  -------------------------------------------
\subsubsection{Models with stellar evolution} 
Figure \ref{fig:FGonly_massloss} shows, on the right, the mass (top) and normalized mass loss (bottom) evolution for the models listed in Table \ref{tab:stev_massloss}, where stellar evolution is taken into account. 
For comparison, the initial conditions we have here adopted are the same as for the models without stellar evolution described above.

As expected, mass removal due to stellar evolution leads to a shallower cluster potential well and, therefore, spurs subsequent mass loss, in the form of lost stars. For models with $W_0=2$, the addition of stellar evolution leads to the dissolution of the cluster after $\sim 3 $ Gyr, due to the initially shallow potential. In all other cases, stellar evolution is less catastrophic, even though the final mass of the cluster is significantly smaller, from one-third to half, with respect to the case without stellar evolution.
 As before, initially segregated clusters suffer a stronger mass loss than not segregated ones.

\subsection{Models with second generation stars}

In Table \ref{tab:SG_massloss}, we report the initial conditions for the thirteen simulations we have performed, taking into account the stellar evolution and with the SG, together with the final values of masses, half-mass radii, the fraction of enriched stars belonging to the final cluster (e.g. considering only bound stars), the fraction of unbound SG stars and central density. %We have adopted the values of $M_{i,SG}$, $W_0^{SG}$ and $r_h^{SG}$ in order to resemble the outcomes of the hydrodynamic simulations of \citet{calura2019}  
It has to be stressed that, as for the previous models, the reported initial value for $r_{h,FG}$ is not the half-mass radius at the time of FG formation, but after the gas expulsion and violent relaxation phases, when the system is considerably more extended than at its formation. During these phases, the half-mass radius of a cluster can increase by a factor of 3 or 4 \citep{lada1984,baumgardt2007}. Its exact value depends on many parameters, e.g. the IMF, the star formation efficiency, the gas and stellar density and the gas expulsion timescale. Such large initial radii are also confirmed by observations of star-forming clusters at high-redshift, where systems extending for several tens of parsecs have been detected. Further discussion regarding the scale radius of star-forming stellar clusters is reported in Section \ref{sec:scalerad}.

In all models, we have assumed that the SG is initially more centrally concentrated than the FG and, at the end of the simulations, all clusters still show, at different degrees, this configuration.

%Once mass segregation is assumed, the FG half mass radius moves from 37pc to 43pc
We have varied several parameters in order to determine their effects on the evolution of the system focusing on the fraction 
of SG stars, which has been determined for several GCs, both in the Milky Way and in external galaxies \citep{milone2020b,dondoglio2021}.
We here calculate it as the mass of bound SG particles over the mass of all the bound particles, $f_{\rm enriched}(<r)=M_{SG}(<r)/M_{tot}(<r)$, both within the half-mass radius and for the entire cluster. In Section \ref{sec:bestmodel}, we discuss the caveats that need to be considered when comparing the theoretical value to the observed one.%its value cannot always be directly compared with the observed one.

In all models, clusters start with the same FG mass and given the possible non-self-similarity of the SG formation (e.g. different initial SG fraction for clusters of different initial mass, see \citealt{yaghoobi2022}), we do not study here the trend of $f_{\rm enriched}$ with cluster mass, which will be addressed in a future work.

First, we studied models with different values of the central velocity dispersion, equal to ${\rm 0\ km\ s^{-1}}$, ${\rm 10\ km \ s^{-1}}$ and the velocity dispersion of the generated King model. We find that such a quantity weakly affects the evolution, and the resulting clusters possess very similar fractions of SG stars. For this reason, all the subsequent simulations have been performed assuming the same velocity dispersion distribution, corresponding to the King one.

As found for the previous models, with and without stellar evolution and without the second generation, a segregated cluster loses a larger amount of mass than not segregated ones. 
Comparing models ${\rm 5N7M3.7K}$ and ${\rm 5Y7M3.7K}$, we derive that the FG loses significantly more mass in the segregated system, while the SG mass is weakly affected, since we have imposed the segregation only to the FG. As a consequence, the $f_{\rm enriched}$ is higher for the segregated system, especially within the half-mass radius.

We have then varied the $W_{0,SG}$ parameter, but we found very weak effects on the evolution of the system, both in terms of mass loss by the two populations, $f_{\rm enriched }$, half-mass radii and central density (see the pairs ${\rm 5Y7M1K-5Y5M1K}$, ${\rm 2N7m1K-2N5m1K}$, ${\rm 2N7m6K-2N5m6K}$).

The parameters whose variation has a stronger and more complex impact on the mass loss and $f_{\rm enriched}$ are, instead, $W_{0,FG}$, $M^{ini}_{SG}$ and the initial $r_{h,SG}$. In general, analogously to what has been derived for the simulations without SG, the greater the concentration of the FG, and therefore the larger the $W_{0,FG}$ value, the less the mass lost by the whole cluster, as well as by the two populations separately, at the end of the simulation. 
While for $W_{0,FG}=5$, a low mass of the SG equal to $7\times 10^5 {\rm M_{\odot}}$ (model ${\rm 5Y7m1K}$) leads to a mild FG mass loss and a small final $f_{\rm enriched}$ (0.16), for $W_{0,FG}=2$, the same initial SG mass (e.g. model ${\rm 2N7m1K}$) allows losing more than $95\%$ of the FG mass. Consequently, when $W_{0,FG}=2$, the final FG mass is one order of magnitude lower than in the model with $W_{0,FG}=5$,{ and the cluster reaches a final $f_{\rm enriched}$ of 0.59 slightly lower than the typical values ($\sim 0.6-0.8$, see \citealt{milone2017} and \citealt{dondoglio2021}) observed in GCs of the same mass.}
Similarly, model ${\rm 2N7M1K}$ loses more FG mass than ${\rm 5Y7M1K}$, reaching a final SG fraction of 0.58.

\begin{figure*}
   \centering
   \includegraphics[width=1\textwidth]{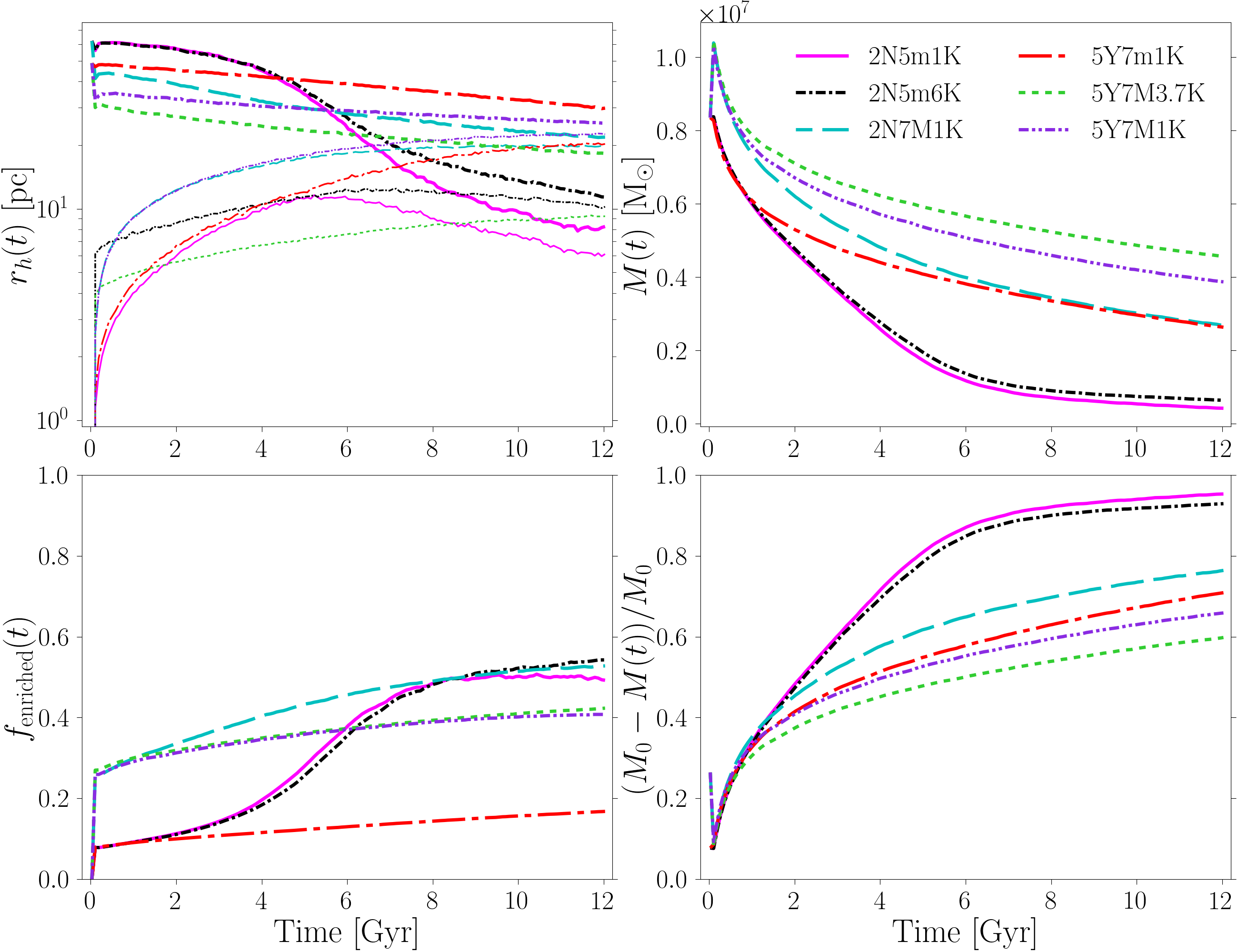}

   \caption{Evolution of the half mass radius (top left), mass (top right), SG mass fraction of the whole cluster (bottom left), mass loss (bottom right) for some of the models in Table \ref{tab:stev_massloss} reported in the legend. The thick lines in the top panels represent the whole cluster, and the thin the SG.}
              \label{fig:SG_massloss}%
    \end{figure*}

Interestingly, however, for fixed $W_{0,FG}=2$, variations of $M^{ini}_{SG}$ lead to significantly different final masses and $r_{h,tot}$, but $f_{\rm enriched}$ slightly changes (models ${\rm 2N7M1K-2N7m1K}$). In model ${\rm 2N7M1K}$, a higher SG mass implies a larger initial SG fraction, so even though the cluster has lost almost an order of magnitude less mass, and consequently has a cluster radius more than double, the final SG fraction is very similar to the one of model ${\rm 2N7m1K}$. On the other hand, for fixed $W_{0,FG}=5$, variations of $M^{ini}_{SG}$ lead to similar FG masses but significantly different SG ones (see $\rm 5Y7M1K- 5Y5m1K$). Therefore, the $f_{\rm enriched}$ values differ by almost a factor of 2. 
These differences in the evolution suggest that there is not a positive correlation between the initial and final values of $f_{\rm enriched}$. This is also visible in Figure \ref{fig:SG_massloss}, where models assuming the same $W_{0,FG}$ do not always follow the same evolution. Therefore, clusters with an initially higher SG fraction do not straightforwardly have a higher final one.

% For a large SG initial mass equal to $3\times 10^6 {\rm M_{\odot}}$, the mass loss of the FG is not that heavily affected by the change of $W_{0,FG}$ implying a significantly smaller variation of the  SG fraction. Therefore, in a shallower FG, a low mass SG permits to lose a significant amount of FG stars and consequently increase the final fraction of SG stars.

A similar behaviour can be found when changing the initial $r_{h,SG}$. While for clusters with a massive initial SG, a smaller $r_{h,SG}$ leads to a stronger SG expansion and lower $f_{\rm enriched}$ (see models $\rm 5Y7M3.7K-5Y7M1K$), for initial low mass SG, clusters with smaller initial $r_{h,SG}$ are more compact at the end of the simulations, with final SG fractions similar to the ones of models starting with larger $r_{h,SG}$ (see models $\rm 2N7m6K-2N7m1K$ and $\rm 2N5m6K-2N5m1K$). Therefore, an initially more compact SG does not imply a lower SG mass loss.

%\textcolor{red}{questa espansione pero e' legata al fatto che inizialmente le due gen non sono in equilibria a causa della segregazione? quindi piu' sono diverse piu' tenderanno a mescolarsi velocemente? }

%\textcolor{red}{!!!!!!!!!versione precendente!!!! For small initial values of $r_{h,SG}$, the SG suffers a significant expansion; for massive SG this leads to final $r_{h,SG}$ larger than the ones obtained with the same models but starting with greater SG half-mass radius (such as $5Y7M1K$ and $5Y7M3.7K$). In addition, $r_{h}$ between the two populations is much more similar than in the case where the initial $r_{h,SG}$ is larger. \textcolor{red}{faster mixing?} For low mass SG clusters, the SG experiences a larger expansion too when the initial $r_{h,SG}$ is smaller, but, after 6 Gyr, the $r_{h,SG}$ starts decreasing (see $2N5m1K$ compared to $2N5m6K$ in \hyperref[fig:SG_massloss]{Figure \ref{fig:SG_massloss}}). The difference in $r_h$ between FG and SG is smaller than for high mass SG, and clusters with smaller initial $r_{h,SG}$ are more compact at the end of the simulation.} %hmodels before being comparable with the FG one and therefore the two generations are characterised by $r_{h,SG}$ half of the FG one. \textcolor{red}{core-collapse?}  

Comparing more quantitatively all the models with observed GCs, with a particular focus on the final masses and half mass radii, it is clearly visible that models assuming $M^{ini}_{SG}=3\times 10^6 {\rm M_{ \odot}}$ and $W_{0,FG}=5$ produce a cluster with a mass almost two times greater than $\omega$ Centauri, the most massive globular cluster known to date \citep{baumgardt2018}. The very massive SG prevents a significant mass loss in these systems, and therefore, also the SG fraction is too small (0.45-0.50) in comparison with the observations. The only exception, in terms of SG fraction, is model ${\rm 5Y7M3.7K}$, where the combination between FG segregation and large initial $r_{h,SG}$ implies a greater loss of FG while poorly affecting the SG. Similar results are obtained decreasing $W_{0,FG}$, like in models ${\rm 2N7M1K}$ and ${\rm 4N7M4.5K}$, where a stronger FG mass loss is taking place, leading to a higher final $f_{\rm enriched}$. Moreover, all models with a massive SG are characterized by a fraction of lost SG stars, with respect to the total unbound mass, $f^{lost}_{SG}$ of $\sim 0.15$. This is between 3 to 5 times larger than what is found for a low-mass SG.

On the other hand, clusters with $M^{ini}_{SG}={7\times 10^5 \rm M_{ \odot}}$ lose much more mass, which depends on the $W_{0,FG}$. As highlighted above, for low $W_{0,FG}$, clusters undergo a strong mass loss, especially in the FG, resulting in final clusters with $M\sim4-6\times 10^5 {\rm M_\odot}$, in agreement with present-day ones. % the higher $W_{0,FG}$, namely the more concentrated the FG, the less mass loss the system undergoes and, in turn, the fraction of SG stars, which initially is of 7\%, is not able to reach the observed values.

\begin{figure*}
        \centering

        \includegraphics[width=1\textwidth]{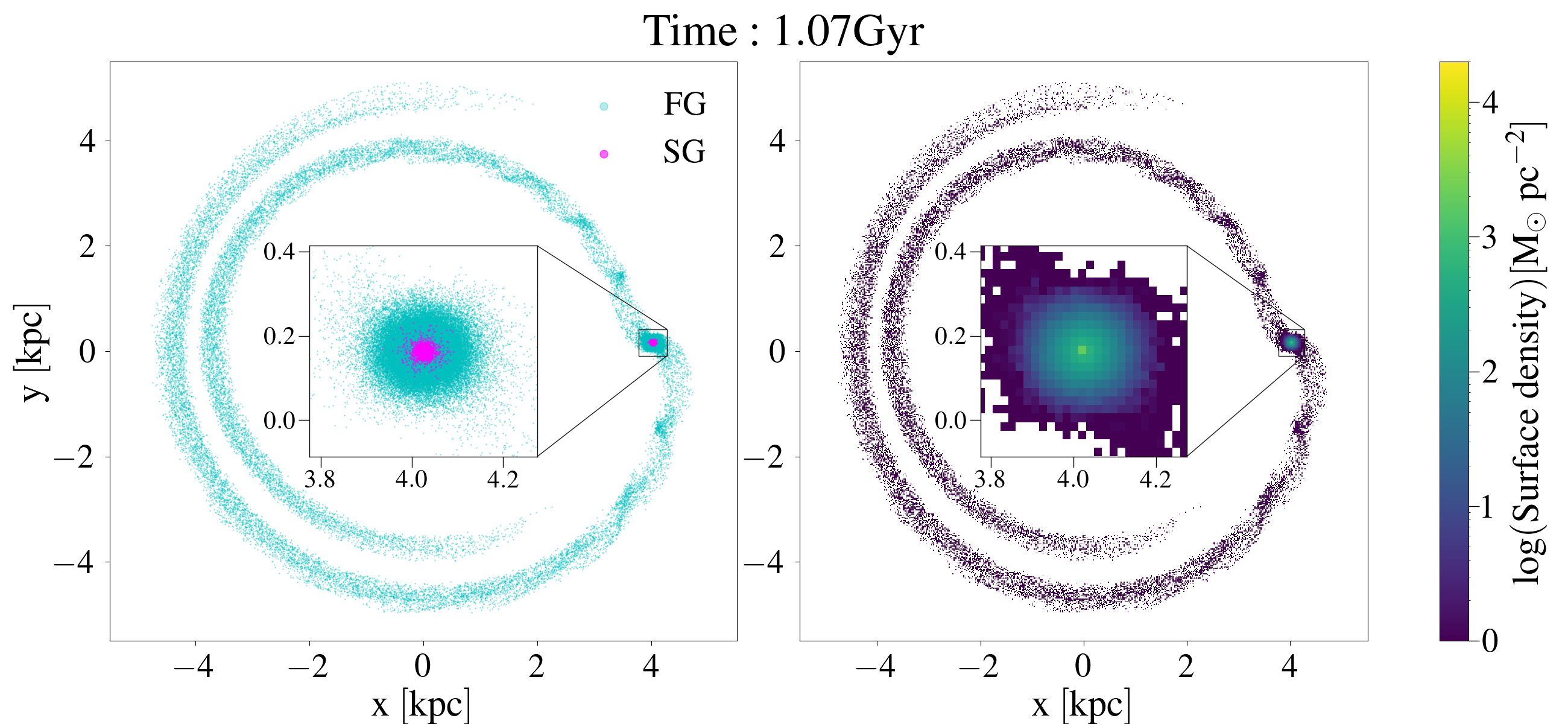}
        \includegraphics[width=1\textwidth]{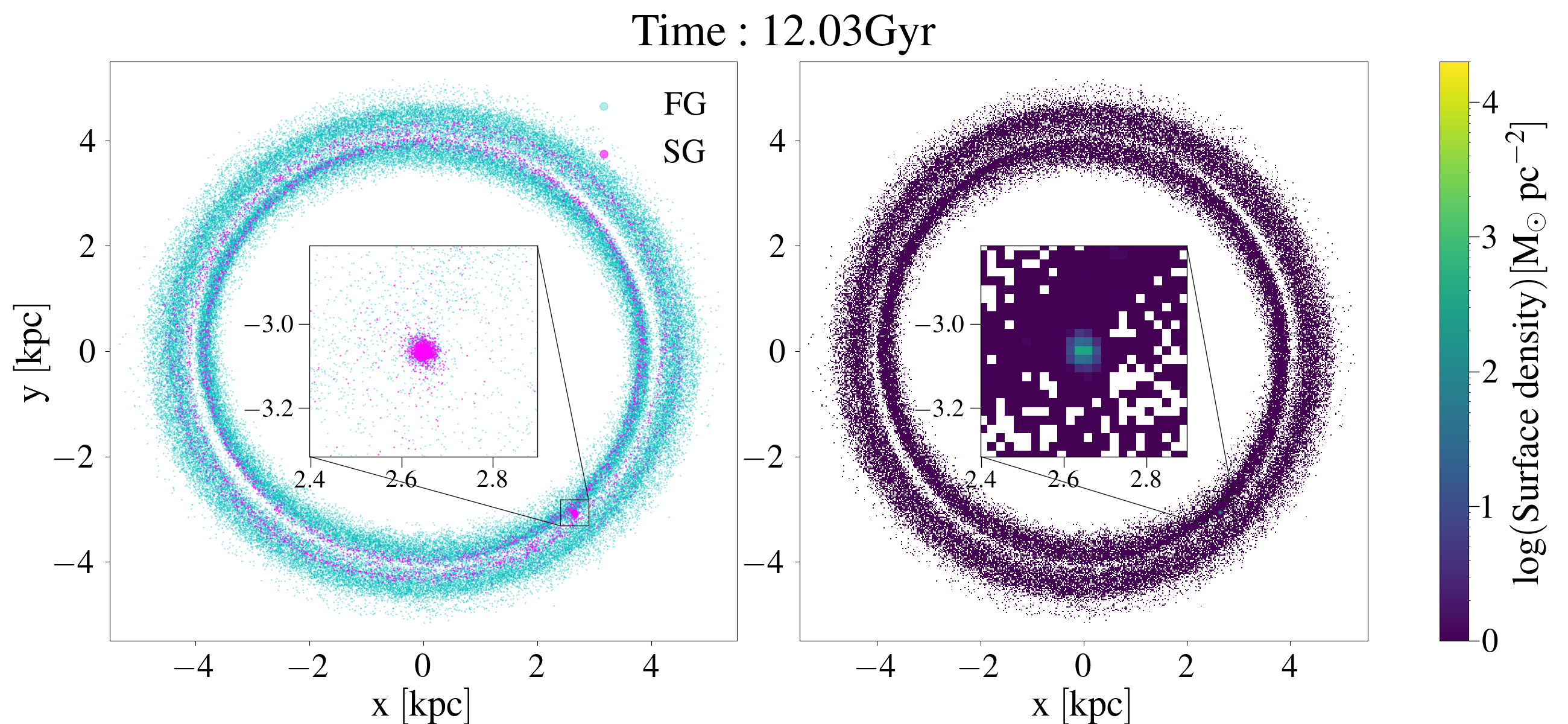}

\caption{Two-dimensional maps at 1 ad 12 Gyr for the model ${\rm 2N7m1K}$. {\it Left panel:} Distribution of the FG (cyan) and SG (magenta). {\it Right panel:} Mass surface density of the whole cluster. At the centre of both panels, we show a zoom-in focused on the centre of mass of the cluster.}
  \label{fig:map7e5}
\end{figure*}

In Figure \ref{fig:SG_massloss} we show the evolution of the models ${\rm 2N5m1K}$, ${\rm 2N5m6K}$, ${\rm 2N7M1K}$, ${\rm 5Y7m1K}$, ${\rm 5Y7M3.7K}$ and ${\rm 5Y7M1K}$ to highlight the effects of changing $W_{0,FG}$, $M^{ini}_{SG}$ and $r_{h,SG}$. Models ${\rm 2N5m1K}$ and ${\rm 2N5m6K}$ are the ones undergoing the strongest mass loss and consequently also their $r_{h}$ and $f_{\rm enriched}$ suffer deep changes during the evolution in the opposite direction; while $r_{h,tot}$ decreases of about 6 times after 12 Gyr, $f_{\rm enriched}$ increases of almost the same amount.

%The model ${\rm 7SG2N5m1K}$, which is the one that better reproduces the SG fraction, is also the one that loses the largest amount of mass equal to 94\% of the whole cluster, out of which 97\% of the FG stars. Moreover, the half mass radius of the cluster drops significantly with respect to the other two models due to the very concentrated SG within an extended FG. Even though model ${\rm 7SG2N7M1K}$ share the same $r_{h,SG}=1$, the formation of a higher mass of SG stars leads to a fast readjustment of the FG distribution, which becomes more concentrated with a half-mass radius of about 40pc. Such a remarkable change of the FG concentration, makes the system more bound and less keen to lose mass. %prevents the loss of FG stars which, in turn, lead to a stronger expansion of the SG.

Interestingly, comparing the results obtained with SG and $W_{0,FG}=2$ with the ones with the same $W_{0,FG}$ but without SG, we clearly see that the formation of a concentrated SG within a shallow FG prevents the disruption of the clusters. An SG, even if not very massive, located at the centre of the system, is enough to strengthen the potential well of the cluster, decreasing the potential energy of the particles which will become more bound.

\subsubsection{Model {\rm  2N7m1K}}
\label{sec:bestmodel}

\begin{figure*}
        \centering

        \includegraphics[width=1\textwidth]{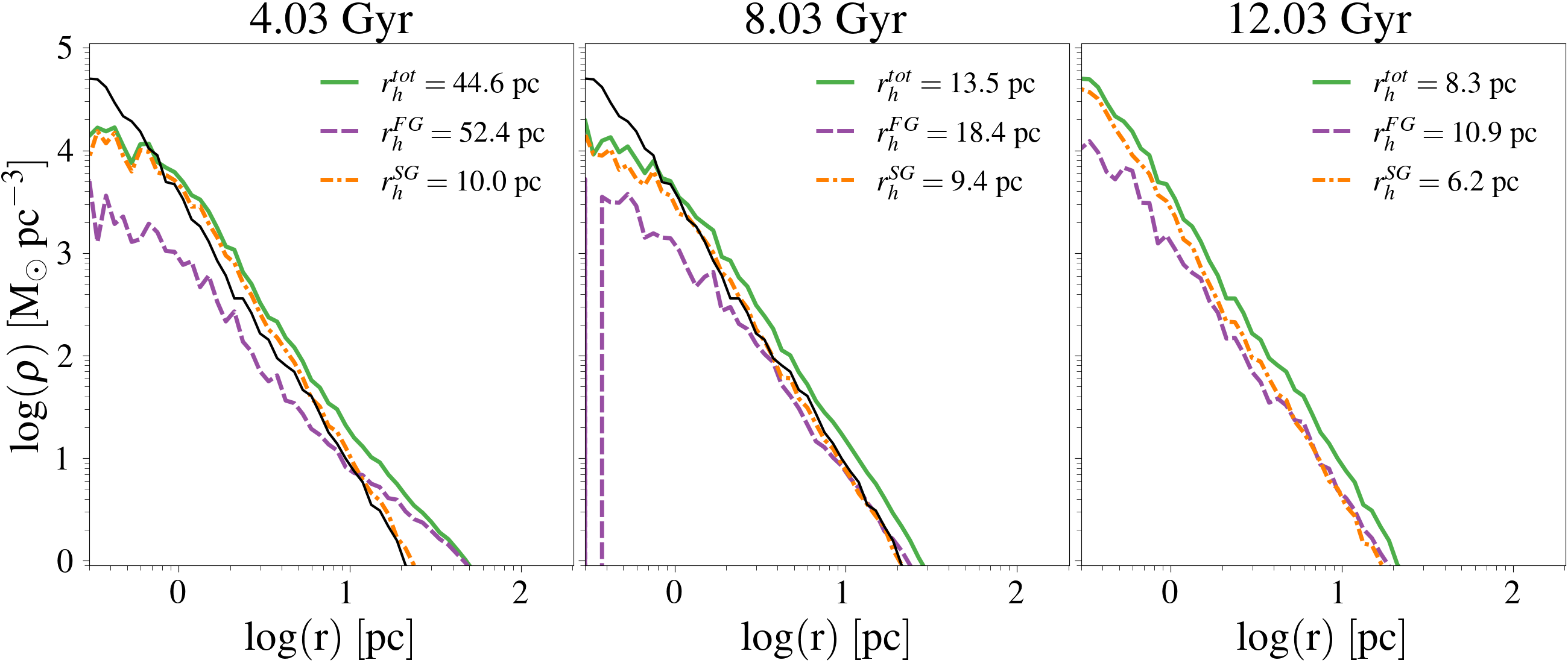}

\caption{Density profiles of the FG (purple, dashed), SG (orange, dash-dotted) and the whole cluster (green, solid) for the model ${\rm 2N7m1K}$ at three different times, indicated on top of each panel. The half-mass radii of the three components are reported in the legend. The density profile at 12 Gyr of the whole cluster is reported in black, in the first two panels, for comparison.}
  \label{fig:densityprof7e5}
\end{figure*}

\begin{figure}
        \centering

        \includegraphics[width=0.5\textwidth,trim={1.2cm 0.cm 0.cm 0cm},clip]{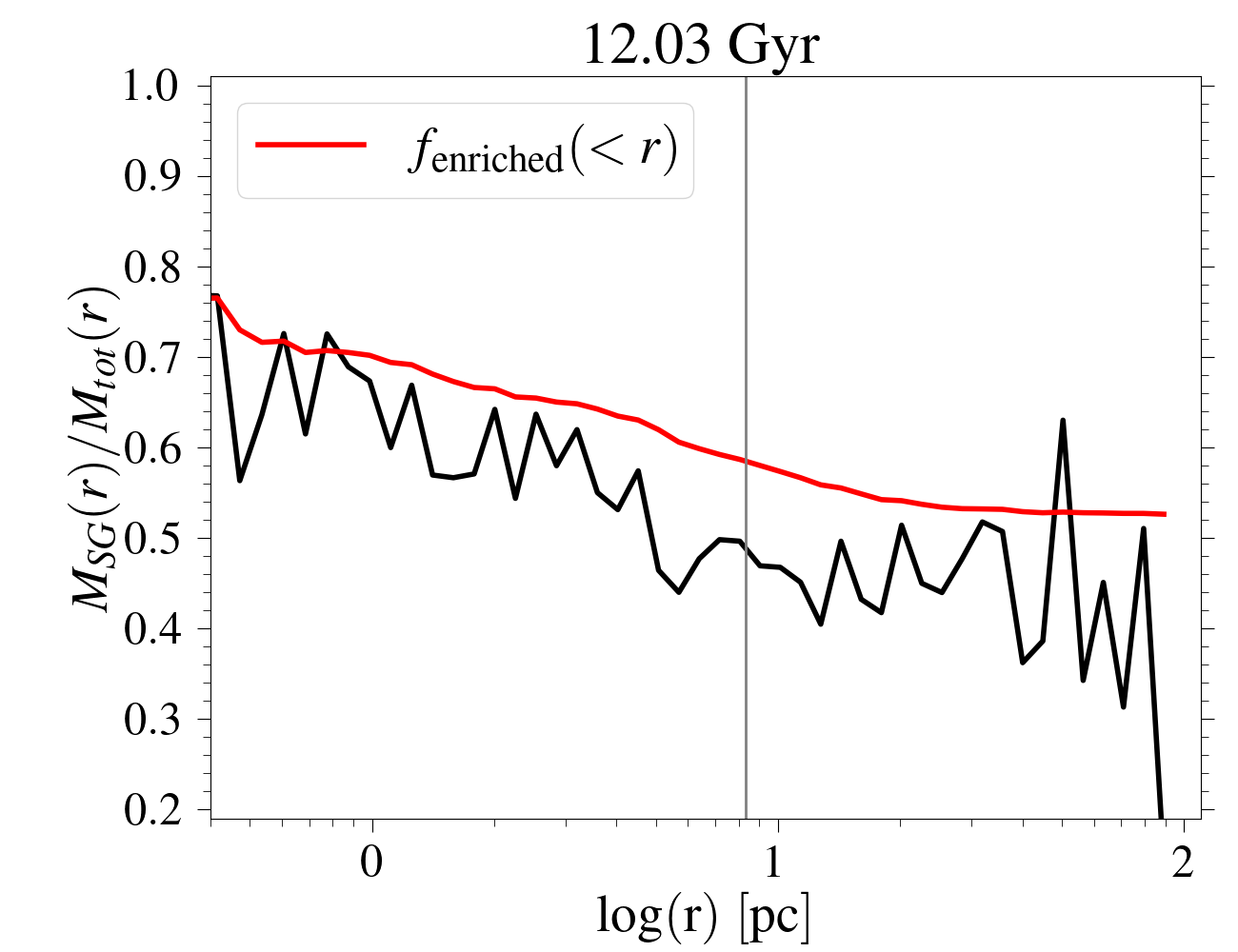}

\caption{ Radial profile of the mass of SG over the total mass $M_{SG}/M_{tot}$ in black and the fraction of enriched stars within $r$, $f_{\rm enriched}$, in red for the model ${\rm 2N7m1K}$ at 12 Gyr. The vertical grey line represents the half-mass radius of the whole cluster, $r_{h,tot}$.}
  \label{fig:cummsg7e5}
\end{figure}

We here focus on the analysis of the model ${\rm 2N7m1K}$, which is the one whose final fraction of SG, defined as $f_{\rm enriched}(r)=M_{SG}(<r)/M_{tot}(<r)$, is more in agreement with current observations, which find fractions between $50\%$ and $80\%$ \citep{milone2017}.

In Figure \ref{fig:map7e5}, we show the distribution of the two populations, on the left, and the projected surface density, on the right, at 1 and 12 Gyr, with a zoom-in centred in the centre of mass of the system in the small inset in the middle of each panel. Already at 1 Gyr, the interaction of the initially spherical cluster with the Milky Way tidal field causes a distortion of the system which develops two significant tidal tails, one leading (internal) and one trailing (external), departing from the centre of the disrupting cluster. Most of the stars in the tails belong to the FG, which, after $12 \Gyr$, is still more extended than the SG, as it was at the beginning of the simulation. Later, the tails lengthen, reaching the main body and two concentric circles appear in the maps. At 12 Gyr, the tails are dominated by the FG, while only 5\% of the particles belong to the SG. Due to the intense mass loss suffered by the FG, the cluster is significantly more compact, as clearly shown in the surface density maps, and dominated by SG stars. Even though the system has been highly distorted, its central region preserves a spherical shape after 12 Gyr.

As expected, the two populations, which were spatially and kinematically different at the beginning, move towards a mixing,  which is spatially highlighted by the change in the density profiles and, in turn, in the half-mass radii. Such behaviour can be clearly seen in Figure  \ref{fig:densityprof7e5}, where we display the density profiles for the two populations together with the one of the whole cluster, compared with the profiles at the end of the simulation. While the SG is always dominant in the centre and its central density does not vary significantly over time, the FG undergoes a notable change in its profile. The FG  increases its central density and decreases its half mass radius, resulting from the loss of stars in the outskirts due to the interaction with the Galactic tidal field. Overall, the central density of the cluster is in good agreement with the ones derived in present-day GCs \citep{baumgardt2018}, while its half-mass radius of $8.3 $ pc is slightly larger than the ones of GCs with mass $\sim 4\times 10^5 {\rm M_{\odot}}$ \citep{mclaughlin2005,Krumholz2019}. It loses almost $98\%$ of FG stars, as generally predicted to match the observed SG fraction, with a final mass loss factor of about 20, which is significantly smaller than the one reported in various other studies.

\begin{figure*}
        \centering

        \includegraphics[width=1\textwidth]{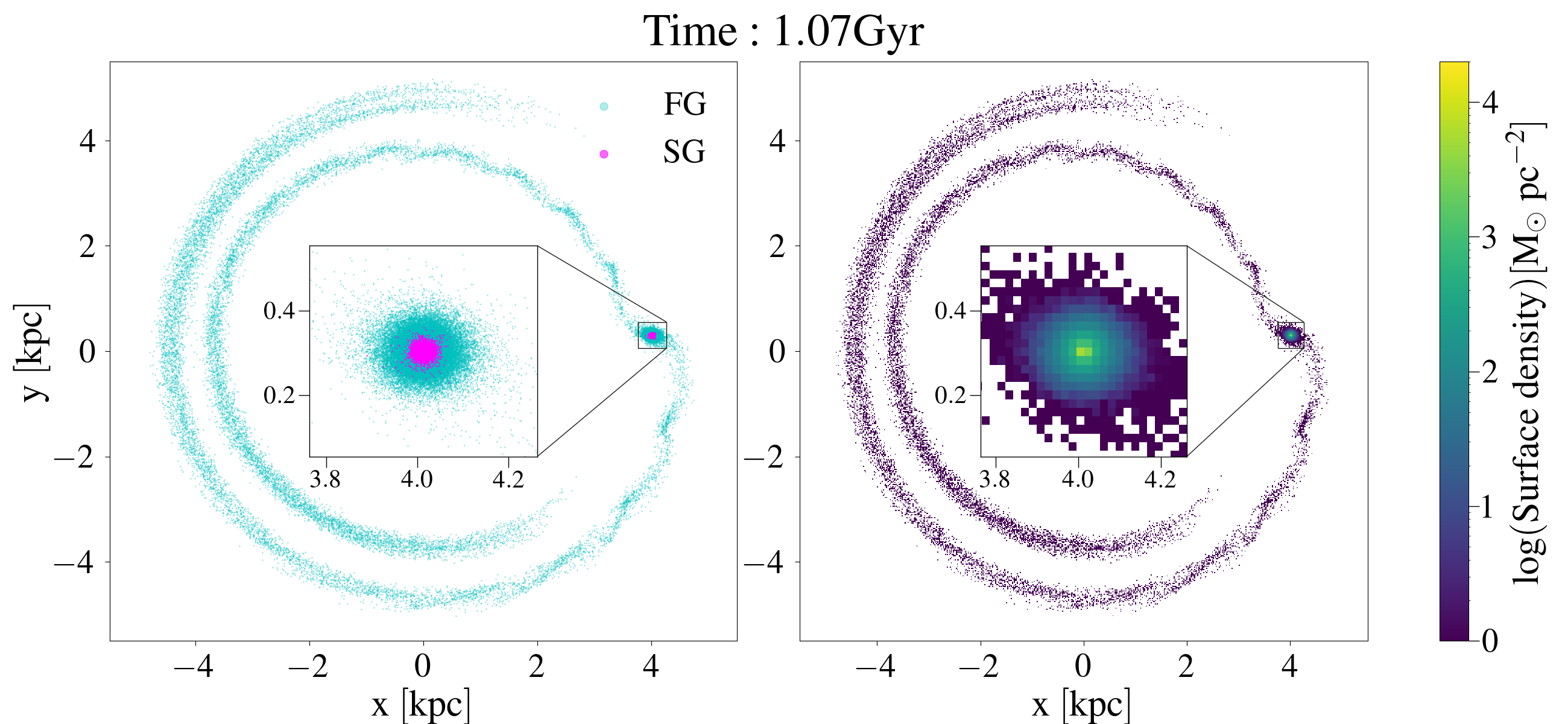}
        \includegraphics[width=1\textwidth]{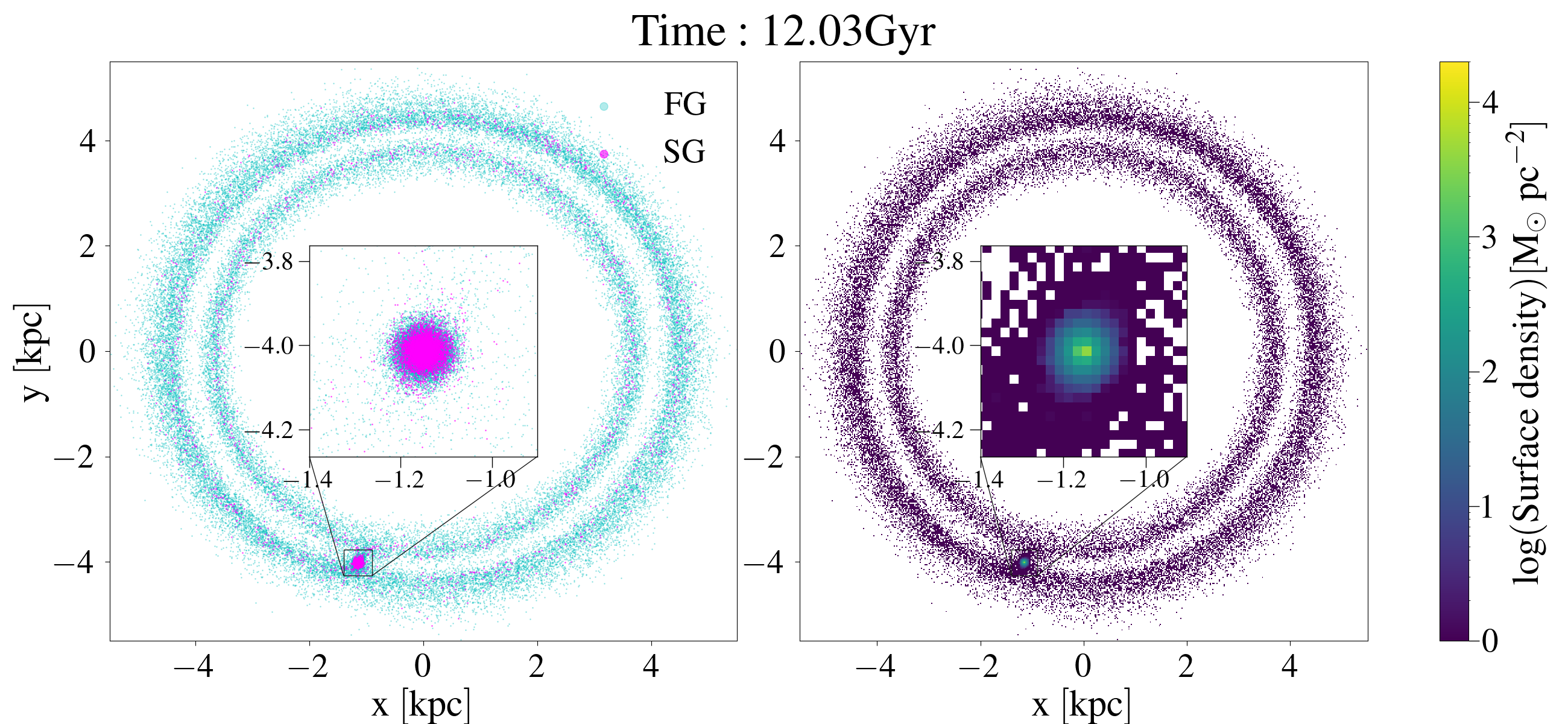}

\caption{Same as in Figure \ref{fig:map7e5} but for model ${\rm 5Y7M3.7K}$.}
  \label{fig:map}
\end{figure*}

Although the two populations become more and more mixed with time, at the end of the simulation they are still well distinguished, with the SG being more concentrated than the FG.
Such difference in the shape of the profiles of the two populations affects the radial fraction of SG stars $M_{SG}/M_{tot}$, which decreases with the distance from the Galactic centre, as shown in Figure \ref{fig:cummsg7e5}. This means that also the fraction of SG stars within a fixed radius $r$, $f_{\rm enriched}(r)$, is not flat all over the cluster, but decreases as well from around 0.7 in the centre, down to 0.54 when considering the whole cluster. Such a decrease has been observed in Milky Way GCs such as 47 Tucanae and NGC 5927 \citep{milone2012b,cordero2014,dondoglio2021,jang2022}, and in simulations \citep{dercole2008}. 
Regarding $f_{\rm enriched}$, it is important here to stress that the observational values of this quantity are rarely calculated for the whole cluster, but, due to the limited field of view (e.g. \citealt{milone2017}), they often refer to the fraction of enriched stars within the inner regions of a cluster (typically between the centre and  $r \sim 0.5-1.5 r_h$).
It is, therefore, important to compute $f_{\rm enriched}$ within regions similar to those observed to account for its possible radial variations.

%. As expected, the initially spherical density distribution of the satellite is distorted by the interaction with the tidal force field of the Milky Way, which produces two significant tidal tails, one leading and one trailing, departing from the main body of the disrupting satellite. However, as illustrated by the zoomed-in surface density maps in the insets in Fig. 3, the central regions remain close to spherical symmetry. 

\subsubsection{Model {\rm  5Y7M3.7K}}

\begin{figure*}
        \centering

        \includegraphics[width=1\textwidth]{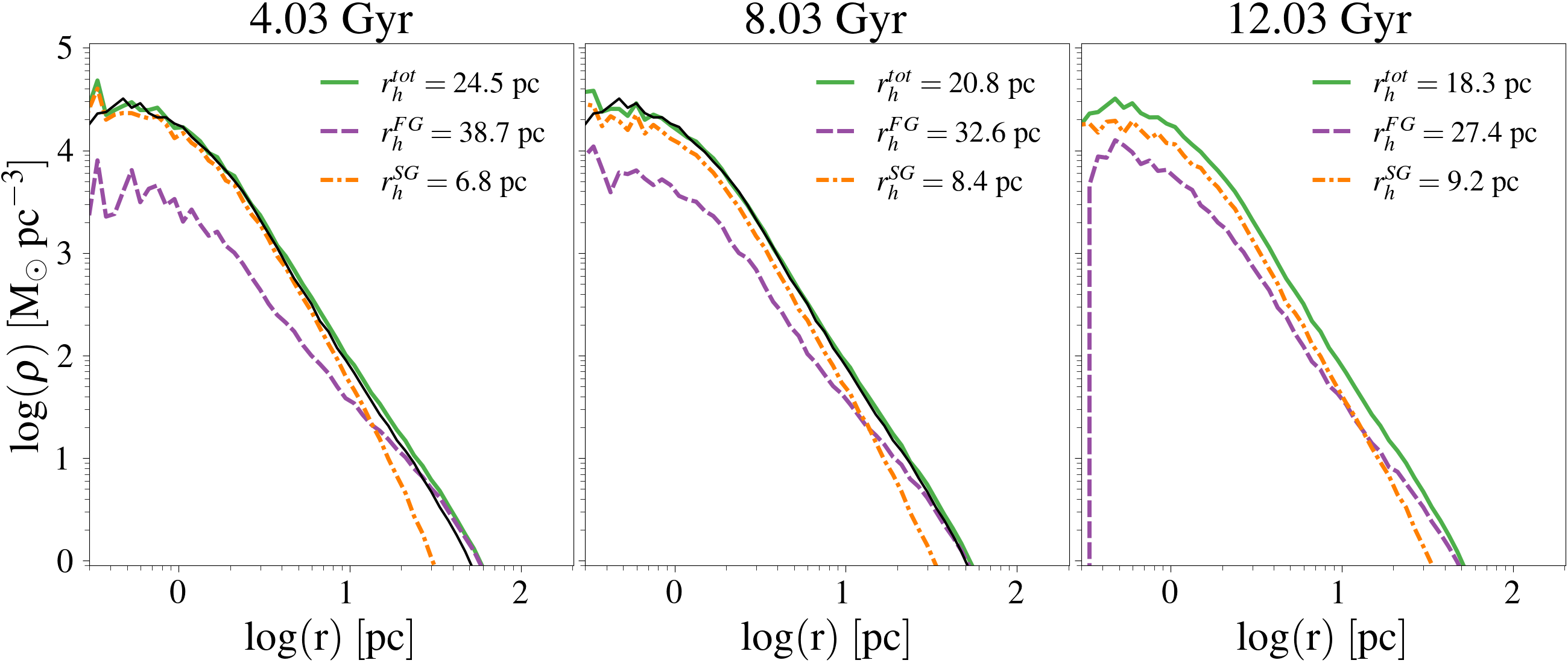}

\caption{Same as in Figure \ref{fig:densityprof7e5} but for model ${\rm 5Y7M3.7K}$.}
  \label{fig:densityprof}
\end{figure*}

\begin{figure}
        \centering

        \includegraphics[width=0.5\textwidth,trim={1.2cm 0.cm 0.cm 0cm},clip]{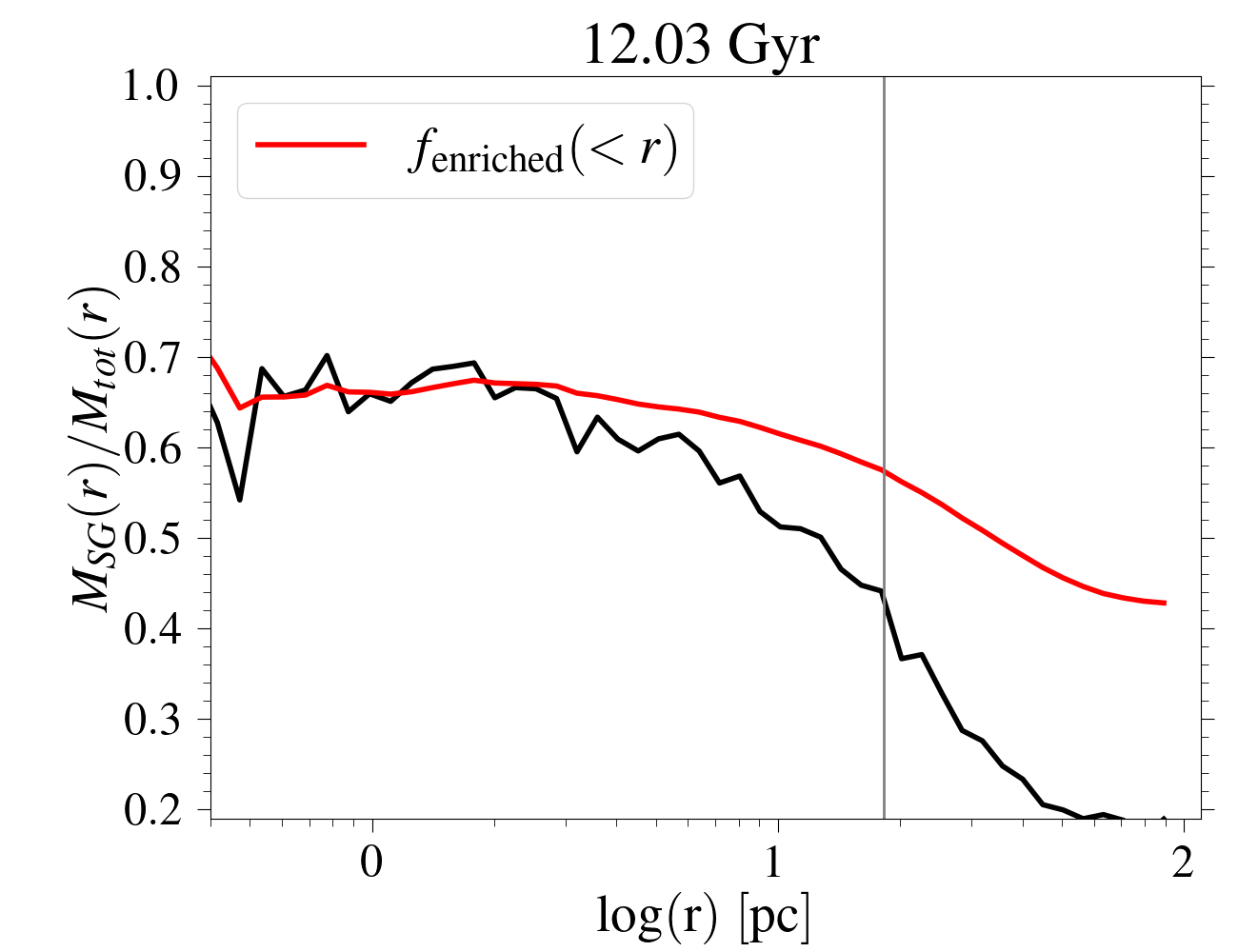}

\caption{ Same as in Figure \ref{fig:cummsg7e5} but for model ${\rm 5Y7M3.7K}$.}
  \label{fig:cummsg}
\end{figure}

For comparison, we report the description of model {\rm 5Y7M3.7K}, whose initial conditions are significantly different from the ones of {\rm 2N7m1K}, but leading to similar $f_{\rm enriched}$.
In Figure \ref{fig:map}, we show the face-on view of the distribution of the two populations, on the left, and of the surface density of the whole cluster, on the right, at two evolutionary times for model {\rm 5Y7M3.7K}. Comparing it with Figure \ref{fig:map7e5} of the model {\rm 2N7m1K}, the central surface density is here slightly higher and, in general, the cluster appears more concentrated at 1 Gyr. At 12 Gyr, the cluster of model {\rm  5Y7M3.7K} is less concentrated, which is reflected in the larger $r_{h}$ reported in Table \ref{tab:SG_massloss}. The cluster has lost fewer stars due to both the higher SG initial mass and the larger $W_{0,FG}$. The milder loss of stars can also be seen looking at the less populated tidal tails.

The profiles of the two models are also quite different; while model {\rm 2N7m1K} has a very steep profile at the centre after ${\rm 12\ Gyr}$ (see Figure \ref{fig:densityprof7e5}), model {\rm  5Y7M3.7K} shows a shallower density profile in the inner regions, as shown in Figure \ref{fig:densityprof}. The density profile of the whole cluster is only slightly changed over time and therefore the final $r_h$ is still very large, not matching the observed values of Galactic GCs. Also, the total mass of the system is significantly greater than the ones of the bulk of Galactic GCs, meaning that with the initial conditions adopted for this model, the cluster is not losing significant mass but still, it ends up with an $f_{\rm enriched }$ in the observed range, as shown in Figure \ref{fig:cummsg}. $f_{\rm enriched}$ is well above 0.6 inside 10 pc while it drops to 0.4 when considering $r>r_{h,tot}$. Here, the difference between the $f_{\rm enriched}$ calculated at the half-mass radius and for the whole cluster is the largest among all models, stressing again the mismatch between the two values.

\section{Discussion}
\label{sec:discussionnb}

Through $N$-body simulations, we have explored the effects of various structural and kinematic properties on the mass loss of a massive GC. We here compare our results with the relevant literature and discuss the strengths and limits of our approach.

\subsection{Comparison with other theoretical works}
Understanding whether a cluster can lose a significant number of FG stars, and therefore reproduce the observational constraints discussed in Section \ref{sec:intro}, has been the goal of several studies in the past.

Firstly, \citet{dercole2008} addressed the issue, proving the feasibility of such a strong mass loss. It is however worth noticing that their simulated cluster has an initial mass of $10^4\ {\rm M_\odot}$, significantly smaller than the one we have adopted here, but also of a typical Galactic GC. Clusters with these masses are more prone to lose mass given their shallower potential well, therefore, it is not surprising that they are able to reach higher values of $f_{\rm enriched}$.

Later, \citet{reinacampos2018}, coupling the EAGLE (Evolution and Assembly of GaLaxies and their Environments; \citealt{crain2015}) simulations with the subgrid model for stellar cluster formation
and evolution MOSAICS (MOdelling Star cluster population Assembly In Cosmological Simulations; \citealt{kruijssen2011}), derived that, once assuming an initial $f_{\rm enriched}=0.05$, their clusters lose a very small number of stars, ending up with fractions of 5-10\% for massive systems. Their result resembles the one we have obtained for the model ${\rm 5Y7m1K}$ which starts with a similar $f_{\rm enriched }$ value. In addition, they stated that extrapolating from their results, a present-day massive cluster with $M_{cl}=10^6 {\rm M_{\odot}}$ should be initially composed of a remarkably high fraction of SG stars ($\sim 0.8$) to reproduce the present-day $f_{\rm enriched}$ values, at variance with what it is generally assumed, and that, even in this case, the cluster would be too extended, with a $r_h\sim 10^2 $ pc, more than one order of magnitude greater than the observed ones. Although in our simulations an extended cluster has to be assumed to match the observed $f_{\rm enriched}$, the final $r_h$ of the whole cluster reduces significantly during the evolution (in the most extreme case of $2N5m1K$ the final cluster $r_h$ is more than 7 times smaller than the FG initial one), which is instead maintained fixed in the \citet{reinacampos2018} study. On the other hand, we noted that a high initial $f_{\rm enriched}$ not always leads to a higher final SG fraction and, therefore, we do not need to assume a massive initial SG to reproduce the observed fraction. Such a mismatch between our results and the ones of \citet{reinacampos2018} may be ascribed to the positive correlation between the initial and final SG fraction assumed in their work, which we have shown is not always true. In addition, the initial $f_{\rm enriched}$ may not be the same for clusters of different masses, as instead assumed in their study, but a positive trend with cluster mass could be imprinted at birth, as shown by \citet{yaghoobi2022}.

Recently, \citet{sollima2022} studied the evolution of multiple stellar populations using the binary fraction as a tool to recover the initial concentration required for the SG. Although we do not include the treatment of binaries, model $2N7m1K$ nicely satisfies the relation required to reproduce the present-day SG binary fraction found by \citet[their eq. 3]{sollima2022}. Similarly, \citet{vesperini2021} and \citet{sollima2021} modelled clusters with masses of the order of $10^6 {\rm M_\odot}$, reproducing the observed SG fractions, much higher than the ones retrieved in this work \citep{hypki2022}. For a thorough comparison, other tests should be performed at lower cluster masses.

%\citep{reinacampos2018,sollima2022,vesperini2013,vesperini2021,baumgardt2018,li2019,rodriguez2022}

\subsection{The fraction of SG lost in the disc}

\begin{comment}

\begin{figure}
        \centering

        \includegraphics[width=0.9\textwidth]{pal5.png}  
        
        \caption{The distribution of stars of the Palomar 5 Galactic globular cluster. The white blob represents the stellar cluster, while the two orange tidal tails are its leading (right) and trailing arms (left), which are extending for almost 10 kpc \citep{odenkirchen2003} and are even more massive than the cluster itself. Credits: Sloan Digital Sky Service.}
  \label{fig:pal5}
\end{figure}
\end{comment}

In Table, \ref{tab:SG_massloss} we list the fraction of unbound SG over the total mass of unbound stars, $f_{SG}^{lost}$. The models can be distinguished into two subgroups: when assuming a massive SG, the unbound stars are composed of $\sim 15\%$ of SG, while low mass SG leads to fractions of $\sim 5\%$. This quantity is important to study GC evolution in terms of mass loss and their contribution to the Galaxy.
Ongoing loss of SG stars has been recently identified both in the tails of the disrupting GC Palomar 5 \citep{phillips2022} and in the bulge cluster NGC 6723 \citep{fernandeztrincado2021b}. Due to the limited sample, a robust fraction of SG in the tails of these two objects cannot be derived, but would be useful to add further constraints on the cluster mass loss.

On the other hand, in the Galactic halo field, several observational studies have searched for SG-like stars identified by their peculiar chemical composition. In particular, recent investigations have determined fractions of $\sim 2-5\%$  \citep{carretta2010c,martell2010,martell2011,ramirez2012,martell2016,koch2019}. However, this is just a lower limit to the fraction of SG stars that are lost by GCs, due to the high uncertainties regarding the fraction of halo stars that were formerly belonging to GCs. Further difficulties arise when focusing on the Galactic disc, where measurements of SG fraction are now not available. With the upcoming arrival of 4MOST and WEAVE coupled with Gaia, new insights into the contribution of GCs to the Galactic disc will be provided.
%Assuming similar fractions between disk and halo, and considering that the fraction of halo stars formerly belonged to GCs are between $10$ (assuming low mass loss) and $50\%$ (assuming that almost $90\%$ of the FG was lost), we conclude that our low mass SG models are slightly overestimating the SG fractions among field stars. On the other hand, the massive SG models predict too many SG, $\sim  30\%$ up to $100\%$, at variance with observations.

\subsection{High-z proto-clusters}

\label{sec:scalerad}
The mass and half-mass radius of present-day clusters are much smaller than the initial values we have assumed for our simulated clusters; however, our assumptions may not be too distant from the real conditions at birth. Accessing the properties of star-forming young GCs is now possible by exploiting gravitational lensing, which permits to revealing faint stellar objects at high redshift, in the epoch of their formation \citep{vanzella2017}. Recently, some proto-GC candidates have been identified, like the ones in the extended star-forming region strongly magnified by the galaxy cluster MACS J0416.1-2403 \citep{vanzella2019,calura2021}. The region is dominated by two star-forming systems: D1, which has a stellar mass of $2.2\times 10^7 {\rm M_\odot}$ and a size of $44\ {\rm pc}$, and T1, which is less massive, with its $2\times 10^6 {\rm M_\odot}$ and a size of $<30\ {\rm pc}$. Interestingly, D1 also shows a nucleated star-forming region surrounded by a diffused component. %These two stellar systems have similar properties to the ones of our simulated clusters, even with the large $r_{h,FG}$ assumed for our best model, suggesting that extended FG may exist in the early Universe.
More extended samples of lensed clumps have been presented \citep{mestric2022,vanzella2022,claeyssens2023},
detected in various lensed fields and across a wide redshift range (from $z\sim 1$ to $z\sim 8$).
These samples are composed of clumps of $10-100$ pc size and mass between $10^5 \Msun$ and $10^9 \Msun$,
therefore including systems with size and mass in the range of our models.

An open problem is to determine if the observed systems represent single star clusters, extended star-forming complexes, super star clusters (SSCs) or dwarf galaxies.  
In the MPs framework, the idea that GCs may form in hierarchical complexes or SSCs is not new \citep{bekki2017}.  
Young GCs might be embedded into a larger structure with similar properties, and a portion of the parent galaxy or SSCs may provide processed materials for the creation of MPs generations \citep{renzini2022}. 

The James Webb Space Telescope is opening a new window on the high redshift observations of GCs.
Besides compact clumps and young proto-GCs at high redshift, a recent, exciting discovery has revealed the presence of quiescent, evolved and massive GCs associated with their host galaxy in the Sparkler system \citep{mowla2022}.
Considering that now we are only in the earliest stages of
calibration of in-flight data from JWST, we have exciting times ahead of us as it is presumable that the current samples may grow rapidly and provide new, fundamental insights on the formation of GCs. 

\subsection{Model limitations}

We have studied the evolution of a massive cluster orbiting the Milky Way to explore whether it can lose mass, as a result of tidal effects of the Galactic potential. We have derived that, in order to reproduce present-day clusters, an initially very extended FG has to be assumed. However, we have not included ingredients that are known to increase the rate of mass loss, such as gravitational and tidal shocks, dynamical friction and the presence of binaries, and dark remnants and do not change the orbital parameters. The addition of these processes would likely increase the mass loss in our simulations and possibly increase the final $f_{\rm enriched}$. On the other hand, the assumed static Galactic potential overestimates the tidal field acting on the GC, which has been shown to be much weaker at early times \citep{renaud2017}. To overcome these limitations, a study of the dynamical evolution of a GC with MPs is to be performed in a fully cosmological context. Attempts to study the early formation of GCs in cosmological simulations are being performed \citep{Kimm2016,ma2020,li2019}, sometimes with resolution high enough to study the feedback of individual stars \citep{calura2022}. 
Although still challenging, it is foreseeable that in the forthcoming future, such tools will allow us also to model MPs and their long-term dynamical evolution.

%%%%%%%%%%%%%%%%%%%%%%%%%%%%%%%%%%%%%%%%%%%%%%%%%%%%%%%%%%%%%%%%

\section{Conclusions}
\label{sec:conclusionsnb}

Most of the scenarios proposed so far for the formation of multiple stellar populations have to deal with the \enquote{mass budget} problem. To overcome it, what it is generally assumed is that clusters were initially more massive, between 5 to 20 times than they are now (see \citealt{bastian2018} and reference therein). As a consequence, during their evolution they must lose a significant amount of mass in terms of stars, to reconcile with the observational values. We have here investigated, through a series of $N$-body simulations, which are the conditions, if any, for a massive cluster with an initial mass of $M\sim 10^7 {\rm M_{\odot}}$, and composed of two different populations, to undergo a significant mass loss during its evolution and end up, after $12 \Gyr$, with structural properties in agreement with the present-day GC ones. %We  explored,  whether a massive cluster with an initial mass of $M\sim 10^7 {\rm M_{\odot}}$, and composed by two different populations, can undergo a significant mass loss during its evolution and end up, after $12 \Gyr$, with structural properties in agreement with the present-day GC ones. 
Our cluster is located in the disc of the Galaxy, and it orbits around the centre at $4 \kpc$. It is therefore evolving under the effect of the tidal field of the Milky Way. We have tested the effects of various parameters on the mass loss and the fraction of SG stars, $f_{\rm enriched}$, one of the strongest observational constraints. 

 Before performing the simulations with two populations, we investigated the evolution of single-population clusters. These results are useful to determine the effects of various parameters and for a comparison with the ones obtained with two populations.

We here summarise the main results of the work:
\begin{itemize}

    \item Our best model $\rm 2N7m1K$, which starts with $W_{0,FG}=2$ and a low mass SG of $M^{ini}_{SG}=7\times 10^5{\rm M_\odot}$, suffers a strong mass loss, particularly in the FG. It predicts a final total mass of $\sim 4\times 10^5{\rm M_\odot}$ in agreement with present-day GCs \citep{baumgardt2018} with $f_{\rm enriched}$ at ${r_{h,tot}}$ of 0.59, which is slightly lower than the average value for clusters with the same mass but comparable with the ones at the lower edge of the observed interval ($\sim 0.6 - 0.8$, \citealt{milone2017}). On the other hand, the  ${r_{h,tot}}=8\pc$ is slightly larger than the ones derived for clusters of similar mass. The FG is reduced by almost $98\%$ of its initial mass and the final mass loss factor is around 20.%, milder than what is often reported in the literature.
    
    \item The parameters that affect the most the mass loss rate and, in general, the evolution of the clusters are the degree of primordial segregation, the FG initial concentration as determined by the initial value of the King dimensionless central potential $W_{0,FG}$, the initial mass of the SG, $M^{ini}_{SG}$, and the initial half-mass radius of the SG, $r_{h,SG}$. In order to lose enough mass, a $W_{0,FG}=2$ and a low mass SG of $M^{ini}_{SG}=7\times 10^5{\rm M_\odot}$ have to be assumed. Clusters with these initial conditions are able to lose more than $90\%$ of their FG mass, as required to solve the mass budget problem. Such a small $W_{0,FG}$ implies an extended FG with $r_h=60\pc$, which is comparable to the size of diffuse star clusters observed in high-redshift star-forming complexes \citep{mestric2022,claeyssens2023}.

    \item From a comparison between the single population models and the two population ones with $W_{0,FG}=2$, it has been shown that the presence of an SG, even if not very massive, prevents the disruption of the system, as it happens when no SG is included. 

    \item Clusters with an initially higher SG mass, and therefore with a higher SG fraction, $f_{\rm enriched}$, are not always showing a higher final SG fraction with respect to clusters starting with a low mass SG. This is particularly true when small values of $W_{0,FG}$ are assumed. Such behaviour suggests that a positive correlation between the initial and final $f_{\rm enriched}$ may not be always verified.

    \item Our clusters are all initially composed of a centrally concentrated SG. This difference between the spatial distribution of the two populations is also found at the end of the simulations for all models. Consequently, $f_{\rm enriched}$ is not flat as a function of radius, and, in particular, it is higher at the centre and decreases moving outwards. Since observations are hardly ever able to derive $f_{\rm enriched}$ for the whole cluster, but rather for some fraction of $r_h$ only, caution has to be made when comparing the simulation results with observed values. In our cases, differences of up to $20\%$ have been found between $f_{\rm enriched}$ of the whole cluster and $f_{\rm enriched}$ at $r_{h,tot}$.

    \item Clusters with low mass SG lose a small fraction of SG stars, generally between $4$ to $5\%$ of all unbound stars in the tails. On the other hand, clusters with initially massive SG lose $15\%$ of SG stars. These values may be used for comparison with GCs where tidal tails have been detected, such as Palomar 5 and NGC 6723.
    
\end{itemize}

Possible follow-ups of the current work could be to expand it, exploring how the intensity of mass loss depends on the initial properties of both the FG and the SG components (e.g. initial masses, Galactocentric distance, galaxy potential). A promising tool to perform a large series of simulations would be to run a single one-component model, and then interpret it a posteriori as a multi-component system, a technique recently applied by \citet{nipoti2021} to a two-component dwarf-galaxy orbiting the Milky Way (see also \citealt{bullock2005,errani2015}).

 Another important further step would be to implement a more sophisticated treatment of stellar evolution, exploiting population synthesis codes such as POSYDON \citep{posydon2022} or SEVN \citep{sevn2019,iorio2023} with the possibility, with the latter, to track the chemical evolution, fundamental for the study of multiple stellar populations.

Moreover, to achieve a more realistic modelisation of the phenomenon, the implementation of other physical processes (e.g. binaries, natal kicks, shocks), not taken into account here, will be crucial, given that could potentially trigger more mass loss.

The models could also be adapted to study external galaxies, such as the Magellanic Clouds, where dynamically younger and FG-dominated globular clusters have been found. Such studies will contribute to refining our knowledge of stellar cluster evolution and the assembly history of the Galaxy.

\bigskip
\textit{ \small Acknowledgements:}
 {\small The authors thank the anonymous referee for a very constructive report and suggestions that helped significantly improve the quality of the manuscript. EL acknowledges financial support
from the European Research Council for the ERC Consolidator
grant DEMOBLACK, under contract no. 770017. This work has received funding from INAF Research GTO-Grant Normal RSN2-1.05.12.05.10 - {\it Understanding the formation of globular clusters with their multiple stellar generations} (ref. Anna F. Marino) of the "Bando INAF per il Finanziamento della Ricerca Fondamentale 2022". Part of the calculations presented in this paper were enabled by resources provided by the Swedish National Infrastructure for Computing (SNIC) at Tetralith and LUNARC. Those resources are partially funded by the Swedish
Research Council through grant agreement no. 2018-05973.
AMB acknowledges funding from the European Union’s Horizon 2020 research and innovation programme under the Marie Skłodowska-Curie grant agreement No 895174. FC acknowledges support from grant PRIN MIUR 2017- 20173ML3WW 001, from the INAF main-stream (1.05.01.86.31) and from PRIN INAF 1.05.01.85.01. }
\\

\bibpunct{(}{)}{;}{a}{}{,} % to follow the A&A style

% for the bibliography, at the end
\bibliographystyle{aa} % style aa.bst
\bibliography{mass_loss}

\end{document}